\documentclass{svjour3}
\journalname{GRG}
\usepackage{graphicx}
\usepackage{rotating}
\usepackage{amsmath}
\usepackage{amssymb}
\usepackage{diagxy}
\usepackage[all]{xy}
\usepackage{eucal}
\newcommand{\reals}{\mathbb{R}}
\newcommand{\integers}{\mathbb{Z}}
\newcommand{\complex}{\mathbb{C}}
\newcommand{\id}{\mathrm{id}}
\newcommand{\tr}{\mathrm{Tr}}
\newcommand{\grad}{\mathrm{grad}}
\newcommand{\EmbSpace}{\mathrm{Emb}(\Sigma{,}M)}
\newcommand{\Emb}{\mathcal{E}}
\newcommand{\Diff}{\mathrm{Diff}(\Sigma)}
\newcommand{\DiffFConn}{\mathrm{Diff}_{\mathrm{F}}^0(\Sigma)}
\newcommand{\DiffF}{\mathrm{Diff_F}(\Sigma)}
\newcommand{\diffF}{\mathfrak{diff}_{\mathrm{F}}(\Sigma)}
\newcommand{\Riem}{\mathrm{Riem}(\Sigma)}
\newcommand{\Conf}{\mathrm{C}(\Sigma)}
\newcommand{\Hor}{\mathrm{Hor}}
\newcommand{\hor}{\mathrm{hor}}
\newcommand{\Ver}{\mathrm{Vert}}
\newcommand{\MCGF}{\mathrm{MCG}_{\mathrm{F}}(\Sigma)}
\newcommand{\Frame}{\mathrm{F}(\Sigma)}

\newcommand{\Ric}{\mathrm{Ric}}

\newcommand{\Sup}{\mathcal{S}(\Sigma)}
\newcommand{\ConfSup}{\mathcal{CS}(\Sigma)}
\newcommand{\SupF}{\mathcal{S}_{\mathrm{F}}(\Sigma)}
\newcommand{\ConfSupF}{\mathcal{CS}_{\mathrm{F}}(\Sigma)}
\newcommand{\G}{\mathcal{G}}
\newcommand{\Hcal}{\mathcal{H}}
\newcommand{\Dcal}{\mathcal{D}}

\begin{document}
\title{The Superspace of Geometrodynamics}
\dedication{I dedicate this contribution to the scientific legacy of John Wheeler}
\author{Domenico Giulini}
\authorrunning{Giulini}
\institute{\at %
Max-Planck-Institute for Gravitational Physics
(Albert-Einstein-Institute),\\ 
Am M\"uhlenberg 1, 
D-14476 Potsdam OT Golm,
Germany.\\
\email{domenico.giulini@aei.mpg.de}}

\date{Received: date / Accepted: date}

\maketitle

\begin{abstract}
Wheeler's Superspace is the arena in which Geometrodynamics takes 
place. I review some aspects of its geometrical and topological 
structure that Wheeler urged us to take seriously in the context 
of canonical quantum gravity. 
\end{abstract}

\begin{quote}
``The stage on which the space of the Universe moves is certainly not
space itself. Nobody can be a stage for himself; he has to have a 
larger arena in which to move. The arena in which space does its 
changing is not even the space-time of Einstein, for space-time is 
the history of space changing with time. The arena must be a larger 
object: \emph{superspace}$ \dots$ It is not endowed with three or 
four dimensions---it's endowed with an \emph{infinite} number of 
dimensions.'' (J.A. Wheeler: \emph{Superspace}, Harper's Magazine, 
July 1974, p.\,9) 
\end{quote} 

\section{Introduction} 
\label{sec:Intro}
From somewhere in the 1950's on, John Wheeler repeatedly urged 
people who were interested in the quantum-gravity programme to 
understand the structure of a mathematical object that he 
called \emph{Superspace}~\cite{Wheeler:1968}%
\cite{Wheeler:EinsteinsVision}. The intended meaning of `Superspace' 
was that of a set, denoted by $\Sup$,  whose points faithfully 
correspond to all possible Riemannian geometries on a given 
3-manifold $\Sigma$. Hence, in fact, there are infinitely many 
Superspaces, one for each 3-manifold $\Sigma$. The physical significance 
of this concept is suggested by the dynamical picture of General
Relativity (henceforth abbreviated by GR), according to which 
spacetime is the history (time evolution) of space.
Accordingly, in Hamiltonian GR, Superspace plays the r\^ole of 
the configuration space the cotangent bundle of which gives
the phase space of 3d-diffeomorphism reduced states. Moreover, 
in Canonical Quantum Gravity (henceforth abbreviated by CQG), 
Superspace plays the r\^ole of the domain for the wave function 
which is still subject to the infamous Wheeler-DeWitt equation. 
In fact, Bryce De\,Witt characterised the motivation for his 
seminal paper on CQG as follows:
\begin{quote}
``The present paper is the direct outcome of conversations
with Wheeler, during which one fundamental question in 
particular kept recurring: \emph{What is the structure of the domain
manifold for the quantum-mechanical state functional?''
(\cite{DeWittQTGI:1967}, p.\,115)}
\end{quote}

More than 41 years after DeWitt's important contribution 
I simply wish to give a small overview over some of the 
answers given so far to the question: \emph{What is the structure of 
Superspace?} Here I interpret `structure' more concretely as 
`metric structure' and `topological structure'. But before 
answers can be attempted, we need to define the object at 
hand. This will be done in the next section; and before doing 
that, we wish to say a few more words on the overall motivation. 

Minkowski space is the stage for relativistic particle physics. 
It comes equipped with some structure (topological, affine, 
causal, metric) that is \emph{not} subject to dynamical changes. 
Likewise, as was emphasised by Wheeler, the arena for 
Geometrodynamics is Superspace, which also comes equipped with 
certain non-dynamical structures.
The topological and geometric structures of Superspace are 
as much a background for GR as the Minkowski space is for 
relativistic particle physics. Now, Quantum Field Theory has 
much to do with the automorphism group of Minkowski space
and, in particular, its representation theory. For example, 
all the linear relativistic wave equations (Klein-Gordan, Weyl, 
Dirac, Maxwell, Proca, Rarita-Schwinger, Dirac-Bargmann, etc.)
can be understood in this group-theoretic fashion, namely as 
projection conditions onto irreducible subspaces in some 
auxiliary Hilbert space. (In the same spirit a characterisation 
of `classical elementary system' has been given as one whose 
phase space supports a transitive symplectic action of
the Poincar\'e group~\cite{Bacry:1967}\cite{Arens:1971a}.)
This is how we arrive at the classifying meaning of `mass' 
and `spin'. Could it be that Quantum Gravity has likewise 
much to do with the automorphism group of Superspace? Can 
we understand this group in any reasonable sense and what 
has it to do with four dimensional diffeomorphisms? 
If elementary particles \emph{are} unitary irreducible 
representations of the Poincar\'e group, as Wigner once 
urged, what would the `elementary systems' be that 
corresponded to irreducible representations of the 
automorphism group of Superspace?

I do not know any reasonably complete answer to any of these 
questions. But the analogies at least suggests the possibility 
of some progress \emph{if} these structures and their automorphisms 
could be be understood in any depth. This is a difficult task, 
as John Wheeler already foresaw forty years ago: 
\begin{quote}
``Die Struktur des Superraumes entr\"atseln? Kaum in einem Sprung,
und kaum heute!'' (\cite{Wheeler:EinsteinsVision}, p.\,61)
\end{quote} 

Related in spirit is a recent approach in the larger context of 
11-dimensional supergravity (see~\cite{Damour.Nicolai:2007,Damour.etal:2007}
and references therein), 
which is based on the observation that the supergravity dynamics in 
certain truncations corresponds to geodesic motion of a massless spinning 
particle on an $E_{10}$ coset space. Here the Wheeler-DeWitt metric 
(\ref{eq:WDWmetric-c}) appears naturally with the right GR-value
$\lambda=1$, which in our context is the only value compatible 
with 4-dimensional diffeomorphism invariance, as we will discuss. 
This may suggest an interesting relation between $E_{10}$ and 
spacetime diffeomorphisms.

\section{Defining Superspace}
\label{sec:Superspace}
As already said, Superspace $\Sup$ is the set of all Riemannian 
geometries on the 3-manifold $\Sigma$. Here `geometries' means 
`metrics up to diffeomorphisms'. Hence $\Sup$ is identified as 
set of equivalence classes in $\Riem$, the set of all smooth 
($C^\infty$) Riemannian metrics in $\Sigma$ under the equivalence 
relation of being related by a smooth diffeomorphism. In other 
words, the group of all ($C^\infty$) diffeomorphisms, $\Diff$, 
has a natural right action on $\Riem$ via pullback and the orbit 
space is identified with $\Sup$: 
\begin{equation}
\label{eq:DefSuperspace}
\Sup:=\Riem/\Diff\,.
\end{equation}

Let us now refine this definition. First, we shall restrict 
attention to those $\Sigma$ which are connected and closed
(compact without boundary). We note that Einstein's field 
equations by themselves do not exclude any such $\Sigma$. 
To see this, recall the form of the constraints for initial 
data $(h,K)$, where $h\in\Riem$ and $K$ is a symmetric covariant 
2nd rank tensor-field (to become the extrinsic curvature of $\Sigma$ 
in spacetime, once the latter is constructed from the dynamical 
part of Einstein's equations) 
\begin{subequations}
\label{eq:Constraints}
\begin{eqnarray}
\label{eq:Constraints-a}
\Vert K\Vert^2_h-\bigl(\tr_h(K)\bigr)^2
-\bigl(R(h)-2\Lambda\bigr)&=-&(2\kappa)\rho_m\,,\\
\label{eq:Constraints-b}
\mathrm{div}_h\bigl(K-h\,\tr_h(K)\bigr)&=&(c\kappa)\,j\,,
\end{eqnarray}
\end{subequations}
where $\rho_m$ and $j_m$ are the densities of energy and momentum of
matter respectively, $R(h)$ is the Ricci scalar for $h$, and 
$\kappa=8\pi G/c^4$. Now, it is known 
that for any smooth function $f:\Sigma\rightarrow\reals$ which is 
negative somewhere on $\Sigma$ there exists an $h\in\Riem$ so 
that $R(h)=f$~\cite{Kazdan.Warner:1975}. Given that strong result, 
we may easily solve (\ref{eq:Constraints}) for $j=0$ on any 
compact $\Sigma$ as follows: First we make the Ansatz $K=\alpha h$ 
for some constant $\alpha$ and some $h\in\Riem$. This solves 
(\ref{eq:Constraints-b}), whatever $\alpha,h$ will be. Geometrically 
this means that the initial $\Sigma$ will be a totally umbillic 
hypersurface in spacetime. Next we solve (\ref{eq:Constraints-a}) 
by fixing $\alpha$ so that 
$\alpha^2>(\Lambda+\kappa\,\sup_\Sigma(\rho_m))/3$ and then 
choosing $h$ so that $R(h)=2\Lambda+2\kappa\rho_m-6\alpha^2$,
which is possible by the result just cited because the right-hand 
side is negative by construction. This argument can be generalised 
to non-compact manifolds with a finite number of ends and 
asymptotically flat data~\cite{Witt:1986a}. 

Next we refine the definition (\ref{eq:DefSuperspace}), in that 
we restrict the group of diffeomorphisms to the proper subgroup 
of those diffeomorphisms that fix a preferred point, called 
$\infty\in\Sigma$, and the tangent space at this point:
\begin{equation}
\label{eq:DefDiffF}
\DiffF:=\bigl\{\phi\in\Diff\mid 
\phi(\infty)=\infty,\,
\phi_*(\infty)=\id\vert_{T_\infty\Sigma}\bigr\}\,.
\end{equation}

The reason for this is twofold: First,  if one is genuinely 
interested in closed $\Sigma$, the space $\Sup:=\Riem/\Diff$
is not a manifold if $\Sigma$ allows for metrics with non-trivial 
isometry groups (not all $\Sigma$ do; compare footnote\,\ref{foot:DegSym}). 
At those metrics $\Diff$ clearly does not act freely, so that 
the quotient $\Riem/\Diff$ has the structure of 
a stratified manifold with nested sets of strata ordered according 
to the dimension of the isometry groups~\cite{Fischer:1970}. 
In that case there is a natural way to minimally resolve the 
singularities~\cite{Fischer:1986} which amounts to taking instead 
the quotient $\Riem\times\Frame/\Diff$, where $\Frame$ is the 
bundle of linear frames over $\Sigma$. The point here is that 
the action of $\Diff$ is now free since there simply are 
no non-trivial isometries that fix a frame. Indeed, if $\phi$ 
is an isometry fixing some frame, we can use the exponential 
map  and $\phi\circ\exp=\exp\circ\phi_*$ (valid for any isometry) 
to show that the subset of points in $\Sigma$ fixed by $\phi$ 
is open. Since this set is also closed and $\Sigma$ is connected, 
$\phi$ must be the identity. 

Now, the quotient $\Riem\times\Frame/\Diff$ is 
isomorphic\footnote{``Isomorphic as what?'' one may ask. 
The answer is: as ILH (inverse-limit Hilbert) manifolds. 
In the ILH sense the action of $\Diff$ on $\Riem\times\Frame$ 
is smooth, free, and proper; see \cite{Fischer:1986} for 
more details and references.} to  
\begin{equation}
\label{eq:DefSupF}
\SupF:=\Riem/\DiffF\,,
\end{equation}
albeit not in a natural way, since one needs to choose a preferred
point $\infty\in \Sigma$. This may seem somewhat artificial if 
really all points in $\Sigma$ are considered to be equally real, 
but this is irrelevant for us as long as we are only interested in 
the isomorphicity class of Superspace. On the other hand, if we 
consider $\Sigma$ as the one-point compactification of a manifold 
with one end\footnote{The condition of `asymptotic flatness' of an 
end includes the topological condition that the one-point 
compactification is again a manifold. This is the case iff there 
exists a compact subset in the manifold the complement of which 
is homeomorphic to the complement of a closed solid ball in 
$\reals^3$.}, then (\ref{eq:DefSupF}) 
would be the right space to start with anyway since then 
diffeomorphisms have to respect the asymptotic geometry 
in that end, like, e.g., asymptotic flatness. \emph{Therefore, from 
now on, we shall refer to $\SupF$ as defined in (\ref{eq:DefSupF}) 
as Superspace.} In view of the original definition (\ref{eq:DefSuperspace}) 
it is usually called `extended Superspace'~\cite{Fischer:1970}.

Clearly, the move from (\ref{eq:DefSuperspace}) to (\ref{eq:DefSupF})
would have been unnecessary in the closed case if one restricted 
attention to those manifolds $\Sigma$ which do not allow for metrics 
with continuous symmetries, i.e. whose degree of symmetry%
\footnote{\label{foot:DegSym}%
Let $\mathcal{I}(\Sigma,h):=\{\phi\in\Diff\mid\phi^*h=h\}$ be the 
isometry group of $(\Sigma,h)$, then it is well known that 
$\dim\mathcal{I}(\Sigma,h)\leq\tfrac{1}{2}n(n+1)$, where 
$n=\dim \Sigma$. $\mathcal{I}(\Sigma,h)$ 
is compact if $\Sigma$ is compact (see, e.g., Sect.\,5 
of~\cite{Myers.Steenrod:1939}). Conversely, if $\Sigma$ allows for 
an effective action of a compact group $G$ then it clearly allows for 
a metric $h$ on which $G$ acts as isometries (just average any 
Riemannian metric over $G$.)  The degree of symmetry of $\Sigma$, 
denoted by $\mathrm{deg}(\Sigma)$, is defined by 
$\mathrm{deg}(\Sigma):=\sup_{h\in\Riem}\{\dim\mathcal{I}(\Sigma,h)\}$.
For compact $\Sigma$ the degree of symmetry is zero iff $\Sigma$ cannot 
support an action of the circle group $SO(2)$. A list of 3-manifolds 
with $\mathrm{deg}>0$ can be found in \cite{Fischer:1970}
whereas \cite{Fischer.Moncrief:1996} contains a characterisation of 
$\mathrm{deg}=0$ manifolds.}%
is zero. Even though these manifolds are not the `obvious' ones
one tends to think of first, they are, in a sense, `most' 3-manifolds.
On the other hand, in order not to deprive ourselves form the 
possibility of physical idealisations in terms of prescribed exact 
symmetries, we prefer to work with $\SupF$ defined in 
(\ref{eq:DefSupF}) (called `extended superspace' 
in~\cite{Fischer:1970}, as already mentioned).

Let us add a few words on the point-set topology of $\Riem$ and 
$\SupF$. First, $\Riem$ is an open positive convex cone in the 
topological vector space of smooth ($C^\infty$) symmetric covariant 
tensor fields over $\Sigma$. The latter space is a Fr\'echet space,
that is, a locally convex topological vector space that admits 
a translation-invariant metric, $\overline{d}$, inducing its topology 
and with respect to which the space is complete. The metric can be
chosen such that $\Diff$ preserves distances. $\Riem$ inherits 
this metric which makes it a metrisable topological space that is 
also second countable (recall also that metrisability implies 
paracompactness). $\SupF$ is given the quotient topology, 
i.e. the strongest topology in which the projection 
$\Riem\rightarrow\SupF$ is continuous. This projection is
also open since $\DiffF$ acts continuously on $\Riem$. 
A metric $d$ on $\SupF$ is defined by 
\begin{equation}
\label{eq:DefMetricOnSup}
d([h_1],[h_2]):=\sup_{\phi_1,\phi_2\in\DiffF}\,
\overline{d}(\phi^*_1h_1,\phi^*_2h_2)\,,
\end{equation}
which also turns $\SupF$ into a connected (being the continuous 
image of the connected $\Riem$) metrisable and second countable
topological space. Hence $\Riem$ and $\SupF$ are perfectly decent 
connected topological spaces which satisfy the strongest separability 
and countability axioms. For more details we refer to 
\cite{Stern:1967}\cite{Fischer:1986}\cite{Fischer:1970}.

The basic geometric idea is now to regard $\Riem$ as principal 
fibre bundle with structure group $\DiffF$ and quotient $\SupF$:
\begin{equation}
\label{eq:PrinFibBundle}
\DiffF\ \mathop{\longrightarrow}^i\ 
\Riem\ \mathop{\longrightarrow}^p\
\SupF
\end{equation}
where the maps $i$ are the inclusion and projection maps respectively. 
This is made possible by the so-called `slice theorems' (see 
\cite{Ebin:1968}\cite{Fischer:1970}), and the fact that the group 
acts freely and properly. This bundle structure has two far-reaching
consequences regarding the geometry and topology of $\SupF$.
Let us discuss these in turn.

\section{Geometry of Superspace}
\label{sec:GeometrySup}
Elements of the Lie algebra $\diffF$ of $\DiffF$ are vector fields 
on $\Sigma$. For any such vector field $\xi$ on $\Sigma$ there is 
a vector field $V_\xi$ on $\Riem$, called the vertical (or fundamental)
vector field associated to $\xi$, whose value at $h\in\Riem$ is 
just the infinitesimal change in $h$ generated by $\xi$, that is, 
\begin{equation}
\label{eq:VertVF}
V_\xi(h)=-L_\xi h\,,
\end{equation}
where $L_\xi$ denoted the Lie derivative with respect to $\xi$. 
Hence, for each $h\in\Riem$, the map $V(h):\xi\mapsto V_\xi(h)$ is an 
anti-Lie homomorphism (the `anti' being due to the fact that we 
have a \emph{right} action of $\Diff$ on $\Riem$), that is 
$[V_\xi,V_\eta]=-V_{[\xi,\eta]}$, if the Lie structure on $\diffF$ 
is that of ordinary commutators of vector fields. The kernel 
of the map $V(h):\xi\mapsto V_\xi(h)$ consists of the 
finite-dimensional subspace of Killing fields on $(\Sigma,h)$. 
The vertical vectors at $h\in\Riem$ therefore form a linear 
subspace $\Ver_h\subset T_h\Riem$, isomorphic to the vector 
fields on $\Sigma$ modulo the Killing fields on $(\Sigma,h)$.
It is a closed subspace due to the fact that the the operator 
$\xi\mapsto L_\xi h$ is overdetermined elliptic 
(cf.~\cite{Besse:EinsteinManifolds}, Appendices G-I).

The family of ultralocal `metrics' on $\Riem$ is given by  
\begin{equation}
\label{eq:GeneralMetricRiem}
\G_{(\alpha,\lambda)}(k,\ell)=\int_\Sigma d^3x\,\alpha\,\sqrt{\det(h)}\,
\bigl(h^{ab}h^{cd}\,k_{ac}\ell_{bd}
-\lambda\, (h^{ab}k_{ab})( h^{cd}\ell_{cd})\bigr)\,, 
\end{equation}
for each $k,\ell\in T_h\Riem$. Here $\alpha$ is a positive real-valued 
function on $\Sigma$ and $\lambda$ a real number. An almost trivial but 
important observation is that $\Diff$ is an isometry group with respect 
to all $\G_{(\alpha,\lambda)}$. The `metric' picked by GR through the 
bilinear term in the constraint (\ref{eq:Constraints-a}) corresponds 
to $\lambda=1$. The positive real-valued function $\alpha$ is not fixed 
and corresponds to the free choice of a lapse-function. In what follows 
we shall focus attention to $\alpha=1$.

The pointwise bilinear form 
$(k,\ell)\mapsto (h\otimes h)(k,\ell)-\lambda\tr_h(k)\tr_h(\ell)$ in the 
integrand of (\ref{eq:GeneralMetricRiem}) defines a symmetric bilinear 
form on the six-dimensional space of symmetric tensors which is positive 
definite for $\lambda<1/3$, of signature $(1,5)$ for $\lambda>1/3$, and 
degenerate of signature $(0,5)$ for $\lambda=1/3$. It defines a metric 
on the homogeneous space $\mathrm{GL}(3)/\mathrm{O}(3)$, where the 
latter may be identified with the space of euclidean metrics on a 
3-dimensional vector space. Parametrising it by $h_{ab}$, we have
\begin{subequations}
\label{eq:WDWmetric}
\begin{equation}
\label{eq:WDWmetric-a}
G_\lambda=G_{\lambda}^{ab\,cd}dh_{ab}\otimes dh_{cd}
=-\epsilon d\tau\otimes d\tau+\frac{\tau^2}{c^2}
\tr(r^{-1}dr\otimes r^{-1}dr)\,,
\end{equation}
where 
\begin{equation}
\label{eq:WDWmetric-b}
r_{ab}:=[\det(h)]^{-1/3}h_{ab},\ \
\tau:=c\,[\det(h)]^{1/4},\ \
c^2:=16\vert\lambda-1/3\vert,\ \
\epsilon=\mbox{sign}(\lambda-1/3),
\end{equation}
and 
\begin{equation}
\label{eq:WDWmetric-c}
G_{\lambda}^{ab\,cd}=\tfrac{1}{2}\sqrt{\det(h)}
\bigl(h^{ac}h^{bd}+h^{ad}h^{bc}-2\lambda h^{ab}h^{cd}\bigr)\,.
\end{equation}
\end{subequations}
This is a 1+5 -- dimensional warped-product geometry in the standard 
form of `cosmological' models (Lorentzian for $\lambda>1/3$), here 
corresponding to the 1+5 decomposition $\mathrm{GL}(3)/\mathrm{O}(3)\cong
\reals\times\mathrm{SL}(3)/\mathrm{SO}(3)$ with scale factor $\tau/c$ 
and homogeneous Riemannian metric on five-dimensional 'space' 
$\mathrm{SL}(3)/\mathrm{SO}(3)$, given by 
$\tr(r^{-1}dr\otimes r^{-1}dr)=r^{ac}r^{bd}\,dr_{ab}\otimes dr_{cd}$.
$\tau=0$ is a genuine `spacelike' (`cosmological') curvature 
singularity.  An early discussion of this finite-dimensional 
geometry was given by DeWitt~\cite{DeWittQTGI:1967}. We stress
that the Lorentzian nature of the Wheeler-DeWitt metric in GR (i.e. for 
$\lambda=1$) has nothing to do with the Lorentzian nature of 
spacetime, as we will see below from the statement of 
Theorem\,\ref{thm:Uniqueness} and formulae~%
(\ref{eq:HamConstrFunct},\ref{eq:DeWittMetricMomenta}); rather, 
it can be related to the attractivity of gravity~%
\cite{Giulini.Kiefer:1994}. 

As for the infinite-dimensional geometry of $\Riem$, we remark 
that an element $h$ of $\Riem$ is a section in 
$T^*\Sigma\otimes T^*\Sigma$ and so is an element of $T_h\Riem$. 
The latter has the fibre-metric (\ref{eq:WDWmetric}). 
It is sometimes useful to use $h$ (for index raising) in order 
to identify $T_h\Riem$ with sections in 
$T\Sigma\otimes T^*\Sigma\cong\mathrm{End}(T\Sigma)$, also 
because the latter has a natural structure as associative- (and 
hence also Lie-) algebra.  Then the inner product 
(\ref{eq:GeneralMetricRiem}) for $\alpha=1$ just reads
(here and below $d\mu(h)=\sqrt{\det(h)}\,d^3x$)
\begin{equation}
\label{eq:GeneralMetricRiem-Alt}
\G_\lambda(k,\ell)=\int_\Sigma d\mu(h)
\bigl(\tr(k\cdot\ell)-\lambda\,\tr(k)\tr(\ell)\bigr)\,. 
\end{equation}

For $\lambda=0$ the infinite-dimensional geometry of $\Riem$ has 
been studied in \cite{Freed.Groisser:1989}. They showed that 
all curvature components involving one or more pure-trace directions 
vanish and that the curvature tensor for the trace-free directions 
is given by (now making use of the natural Lie-algebra structure of 
$T\Sigma\otimes T^*\Sigma$)
\begin{equation}
\label{eq:CurvTensorTracefree} 
R(k,\ell)m=-\tfrac{1}{4}[[k,\ell],m]\,.
\end{equation}
In particular, this implies that the sectional curvatures
involving pure trace directions vanish and that the sectional 
curvatures for trace-free directions $k,\ell$ are non-positive:
\begin{equation}
\label{eq:SectCurvTracefree}
\begin{split}
K(k,\ell)&=-\tfrac{1}{4}\int_\Sigma d\mu(h)\,
           \tr\bigl(k\cdot R(k,\ell)\ell\bigr)\\
         &=-\tfrac{1}{4}\int_\Sigma d\mu(h)\,
           \tr\bigl([k,\ell]\cdot[\ell,k]\bigr)\leq 0\,.\\
\end{split} 
\end{equation}

Similar results hold for other values of $\lambda$, though
some positivity statements cease to hold for $\lambda>1/3$. 
We keep the generality in the value of $\lambda$ for the moment 
in order to show that the value $\lambda=1$ picked by GR is quite 
special. Since, as already stated, all elements of $\DiffF$ are 
isometries of $\G_\lambda$, it is natural to try to define a 
bundle connection on $\Riem$ by taking the \emph{horizontal subspace} 
$\Hor_h^\lambda$ at each $T_h\Riem$ to be the 
$\G_\lambda$--orthogonal complement to $\Ver_h$, as suggested 
in~\cite{Giulini:1995c}. From (\ref{eq:GeneralMetricRiem}) one 
sees that $k\in T_h\Riem$ is orthogonal to all $L_\xi h$ iff 
\begin{equation}
\label{eq:HorizontalityCond}
(\mathcal{O}_\lambda k)^a:=-\nabla^b(k_{ab}-\lambda h_{ab}h^{cd}k_{cd})=0\,.
\end{equation} 
But note that orthogonality does not imply transversality 
if the metric is indefinite, as for $\lambda=1$. In that 
case the intersection $\Ver_h\cap\Hor_h^\lambda$ may well 
be non trivial, which implies that there is no well defined 
projection map 
\begin{equation}
\label{eq:HorProjMap}
\hor^\lambda_h: T_h\Riem\rightarrow \Hor_h^\lambda\,.
\end{equation}
The definition of this map would be as follows: 
Let $k\in T_h\Riem$, find a vector field $\xi$ on $\Sigma$ 
such that $k-V_\xi$ is horizontal. Then $V_\xi$ is 
the ($\lambda$ dependent) vertical component of $k$ and 
the map $k\mapsto k-V_\xi$ is the ($\lambda$ dependent) 
horizontal projection (\ref{eq:HorProjMap}). When does 
that work? Well, according to (\ref{eq:HorizontalityCond}),
the condition for $k-V_\xi$ to be horizontal for 
given $k$ is equivalent to the following differential
equation for $\xi$:
\begin{equation}
\label{eq:HorizontalProjCond}
D_\lambda\xi:=
\bigl(
\delta d+2(1-\lambda)d\delta-2\Ric
\bigr)\xi=
\mathcal{O}_h k\,.
\end{equation} 
Here we regarded $\xi$ as one-form and $d,\delta$ denote 
the standard exterior differential and co-differential 
($\delta\xi=-\nabla^a\xi_a$) respectively. Moreover, $\Ric$
is the endomorphism on one-forms induced by the Ricci
tensor ($\xi_a\mapsto R^b_a\xi_b$). Note that the right-hand 
side of (\ref{eq:HorizontalProjCond}) is $L^2$-orthogonal 
for all $k$ to precisely the Killing fields. 

The singular nature of the GR value $\lambda=1$ is now 
seen from writing down the principal symbol of the operator 
$D_\lambda$ on the left-hand side of (\ref{eq:HorizontalProjCond}),
\begin{equation}
\label{eq:SymbOpD}
\sigma_\lambda(\zeta)^a_b=\Vert\zeta\Vert
\left(
\delta^a_b+(1-2\lambda)\frac{\zeta^a\zeta_b}{\Vert\zeta\Vert}
\right)\,,
\end{equation}
whose determinant is $\Vert\xi\Vert^62(1-\lambda)$. 
Hence $\sigma_\lambda$ is positive definite for 
$\lambda<1$ and indefinite but still an isomorphism 
for $\lambda >1$. This means that $D_\lambda$ is elliptic
for $\lambda\ne 1$ (strongly elliptic\footnote{We follow
the terminology of Appendix I in \cite{Besse:EinsteinManifolds}.} 
for $\lambda<1$) but fails to be elliptic for precisely
the GR value $\lambda=1$. This implies the possibility 
for the kernel of $D_{\lambda=1}$ to become infinite 
dimensional. 

The last remark has a direct implication as regards the 
intersection of the horizontal and vertical subspaces.
Recall that solutions $\xi$ to $D_\lambda\xi=0$ modulo 
Killing fields (which always solve this equation) 
correspond faithfully to vertical vectors at $h\in\Riem$ 
(via $\xi\mapsto V_\xi(h)$) which are also horizontal. 
Since Killing vectors span at most a finite dimensional 
space, an infinite dimensional intersection
$\Ver_h\cap\Hor_h^\lambda$ would be implied by an 
infinite dimensional kernel of $D_1$.

That this possibility for $\lambda=1$ is actually realised for 
any $\Sigma$ is easy to see: Take a metric $h$ on $\Sigma$ that 
is flat in an open region $U\subset\Sigma$, and consider 
$k\in T_h\Riem$ of the form $k_{ab}=2\nabla_a\nabla_b\phi$,
where $\phi$ is a real-valued function on $\Sigma$
whose support is contained in $U$. Then $k$ is vertical 
since $k=L_\xi h$ for $\xi=\grad\phi$ (a non-zero gradient 
vector-field is never Killing on a compact $\Sigma$), 
and also horizontal since $k$ satisfies 
(\ref{eq:HorizontalityCond}). Such $\xi$ clearly span an 
infinite-dimensional subspace in the kernel of $D_1$. 

On the other hand, for $\lambda=1$ there are also always 
open sets of $h\in\Riem$ (and of $[h]\in\SupF$) for which 
the kernel of $D_1$ is trivial (the kernel clearly depends 
only on the diffeomorphism class $[h]$ of $h$). 
For example, consider metrics with negative definite Ricci
tensor, which exist for any closed $\Sigma$~\cite{Gao.Yao:1986}. 
(Note that Ricci-negative geometries never allow for non-trivial 
Killing fields.) Then it is clear from the definition of 
$D_\lambda$ that it is a positive-definite operator for 
$\lambda\leq 1$. Hence the intersection 
$\Ver_h\cap\Hor_h^\lambda$ is trivial. 

In the latter case it is interesting to observe that $\G_\lambda$ 
restricted to $\Ver_h$ ($h$ Ricci negative) is positive definite, 
since 
\begin{equation}
\label{eq:VerticalMetricRicciFlat}
\G_\lambda\bigl(V_\xi(h),V_\xi(h)\bigr)
=2\int_\Sigma d\mu(h)\,h(\xi,D_\lambda\xi)\,.
\end{equation}
This means that $\G_\lambda$ restricted to the orthogonal 
complement, $\Hor_h^\lambda$, contains infinitely many negative and 
infinitely may positive directions and the same 
$(\infty,\infty)$ -- signature is then directly inherited 
by $T_{[h]}\SupF$ for any Ricci-negative geometry $[h]$.

Far less generic but still interesting examples for 
trivial intersections $\Ver_h\cap\Hor_h^\lambda$ in case 
$\lambda=1$ are given by Einstein metrics with positive 
Einstein constants. Since this condition implies constant 
positive sectional curvature, such metrics only exist on 
manifolds $\Sigma$ with finite fundamental group, so that 
$\Sigma$ must be a spherical space form $S^3/G$, where 
$G$ is a finite subgroup of $SO(4)$ acting freely on $S^3$. 
Also note that the subspace in $\SupF$ of Einstein 
geometries is finite dimensional. 
\begin{figure}[ht]
\centering
\includegraphics[width=0.75\linewidth]{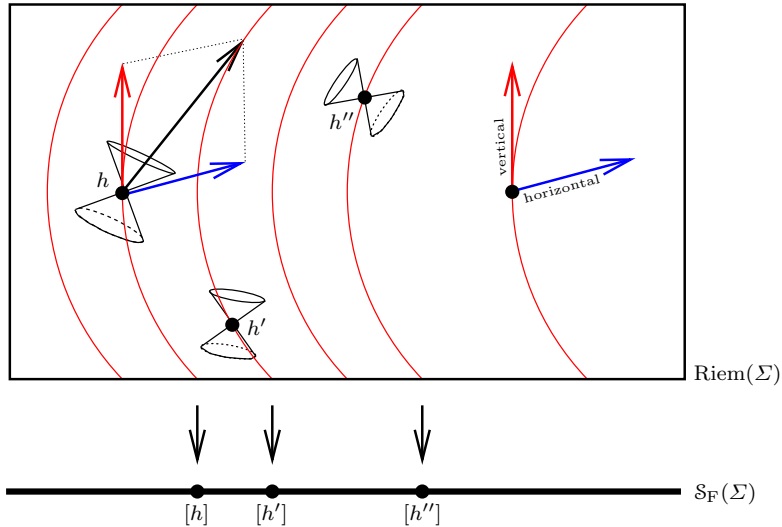}
\put(2,0){\small $\SupF$}
\put(2,45){\small $\Riem$}
\put(-224,118){\small $h$}
\put(-167,62){\small $h'$}
\put(-136,140){\small $h''$}
\put(-191,-8){\small $[h]$}
\put(-164,-8){\small $[h']$}
\put(-108,-8){\small $[h'']$}
\put(-62,112){%
 \begin{rotate}{15}{\tiny horizontal}\end{rotate}}
\put(-69,122){%
 \begin{rotate}{93}{\tiny vertical}\end{rotate}}
\caption{\label{fig:FigGeomSuperspace}%
The rectangle depicts the space $\Riem$ which is 
fibred by the orbits of $\DiffF$ (curved vertical 
lines). The metric $\G_1$ on $\Riem$ is such that 
as we move along $\Riem$ transversal to the fibres 
the ``light-cones'' tilt relative to the fibre 
directions. The process here shows a transition 
at $[h']$ where some fibre directions are lightlike
and no metric can be defined in $T_{[h']}\SupF$, 
whereas they are timelike at $[h]$ and spacelike at 
$[h'']$. The parallelogram at $h$ merely indicates the 
horizontal and vertical components of a vector
in $T_h\Riem$.}
\end{figure}
Now, any solution $\xi$ to $D_1\xi=0$ must be divergenceless 
(take the co-differential $\delta$ of this equation) and 
hence Killing. The last statement follows without computation 
from the fact that $D_1\xi=0$ is nothing but the condition 
that $L_\xi h$ is $\G_1$--orthogonal to all vertical vectors 
in $T_h\Riem$, which for divergenceless $\xi$ (traceless 
$L_\xi h$) is equivalent to $\G_0$--orthogonality, but 
then positive definiteness of $\G_0$ and 
$\G_0(L_\xi h,L_\xi h)=0$ immediately imply $L_\xi h=0$. 
This shows the triviality of $\Ver_h\cap\Hor_h^\lambda$. 

The foregoing shows that $\G_1$ indeed defines a metric 
at $T_{[h]}\SupF$ for Ricci-positive Einstein geometries $[h]$. 
How does the signature of this metric compare to the 
signature $(\infty,\infty)$ at Ricci-negative geometries?
The answer is surprising: Take, e.g., for $[h]$ the 
round geometry on $\Sigma=S^3$. Then it can be shown that
$\G_1$ defines a \emph{Lorentz geometry} on $T_{[h]}\SupF$, that 
is with signature $(1,\infty)$, containing exactly one 
negative direction~\cite{Giulini:1995c}.
This means that the signature of the metric defined at various 
points in Superspace varies strongly, with intermediate 
transition regions where no metric can be defined at all 
due to signature change. Figure\,\ref{fig:FigGeomSuperspace}
is an attempt to picture this situation.

\section{Intermezzo: GR as simplest representation of symmetry}
\label{sec:IntermezzoGRasSymRep}
It is well known that the field equations of GR have certain 
uniqueness properties and can accordingly be `deduced' under 
suitable hypotheses involving a \emph{symmetry principle} 
(diffeomorphism invariance), the equivalence principle, and 
some apparently mild technical hypotheses. More precisely, 
the equivalence principle suggests to only take the metric 
as dynamical variable~\cite{ThorneLeeLightman:1973} 
representing the gravitational field (to which matter then couples 
universally), whereas diffeomorphism invariance, derivability from an 
invariant Lagrangian (alternatively: local energy-momentum conservation
in the sense of covariant divergencelessness), dependence of the 
equations on the metric up to at most second derivatives, and, finally, 
four-dimensionality lead uniquely to the left-hand side of Einstein's 
equation, including a possibly non-vanishing cosmological 
constant~\cite{Lovelock:1972}. Here we will review how this 
`deduction' works in the Hamiltonian setting on phase space 
$T^*\Riem$, which goes back to 
\cite{Hojman.etal:1973,Teitelboim:1973,Kuchar:1973,Hojman.etal:1976}.

\subsection{3+1 decomposition}
Since the 3+1 split of Einstein's equations has already been 
introduced in Claus Kiefer's contribution I can be brief 
on that point. The basic idea is to first imagine a spacetime 
$(M,g)$ being given, where topologically $M$ is a product 
$\reals\times\Sigma$. Spacetime is then considered as the trajectory 
(history) of space in the following way: 
Let $\EmbSpace$ denote the space of smooth spacelike embeddings 
$\Sigma\rightarrow M$. We consider a curve 
$\reals\ni t\rightarrow\Emb_t\in\EmbSpace$ corresponding to 
a one-parameter family of smooth embeddings with spacelike 
images. We assume the images $\Emb_t(\Sigma)=:\Emb_t\subset M$ 
to be mutually disjoint and moreover that 
$\hat\Emb:\reals\times\Sigma\rightarrow M$, $(t,p)\mapsto\Emb_t(p)$, 
is an embedding (it is sometimes found convenient to relax this 
condition, but this is of no importance here). The Lorentz 
manifold $(\reals\times\Sigma,\Emb^*g)$ may now be taken as 
($\Emb$--dependent) representative of $M$ (or at least some open 
part of it) on which the leaves of the above foliation simply 
correspond to the $t=\mathrm{const.}$ hypersurfaces. Let $n$ 
denote a field of normalised timelike vectors normal to these 
leaves. $n$ is unique up to orientation, so that the choice of 
$n$ amounts to picking a `future direction'. 

The tangent vector $d\Emb_t/dt\vert_{t=0}$ at $\Emb_0\in\EmbSpace$  
corresponds to a vector field over $\Emb_0$ (i.e. section in 
$T(M)\vert_{\Emb_0}$), given by 
\begin{equation}
\label{eq:LapseShift}
\frac{d\Emb_t(p)}{dt}\Big\vert_{t=0}
=:\frac{\partial}{\partial t}\Big\vert_{\Emb_0(p)}
=\alpha n+\beta
\end{equation}
with components $(\alpha,\beta)=(\mathrm{lapse},\mathrm{shift})$ normal 
and tangential to $\Sigma_0\subset M$. 

Conversely, each vector field $V$ on $M$ defines a vector field 
$X(V)$ on $\EmbSpace$, corresponding to the left action of
$\mathrm{Diff}(M)$ on $\EmbSpace$ by composition. In local coordinates 
$y^\mu$ on $M$ and $x^k$ on $\Sigma$ it can be written as
\begin{equation}
\label{eq:XofV}
X(V)=\int_\Sigma d^3x\,V^\mu(y(x))\frac{\delta}{\delta y^\mu(x)}\,.
\end{equation}
One easily verifies that $X:V\mapsto X(V)$ is a Lie homomorphism:
\begin{equation}
\label{eq:LieHomo}
\bigl[X(V),X(W)\bigr]=X\bigl([V,W]\bigr)\,.
\end{equation}
In this sense, the Lie algebra of the four-dimensional diffeomorphism
group is implemented on phase space of any generally covariant theory 
whose phase space includes the embedding variables 
\cite{Isham.Kuchar:1985a} (so-called `parametrised theories').

\begin{figure}[ht]
\centering
\includegraphics[width=0.32\linewidth]{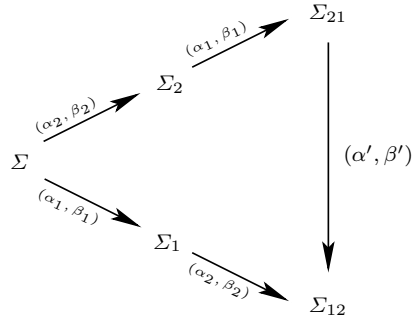}
\put(-123,51){$\Sigma$}
\put(-68,81){$\Sigma_2$}
\put(-69,21){$\Sigma_1$}
\put(-10,108){$\Sigma_{21}$}
\put(-10,-3){$\Sigma_{12}$}
\put(-113,65){\tiny
 \begin{rotate}{25}$(\alpha_2,\beta_2)$\end{rotate}}
\put(-56,94){\tiny
 \begin{rotate}{25}$(\alpha_1,\beta_1)$\end{rotate}}
\put(-113,41){\tiny 
 \begin{rotate}{-25}$(\alpha_1,\beta_1)$\end{rotate}}
\put(-56,13){\tiny 
 \begin{rotate}{-25}$(\alpha_2,\beta_2)$\end{rotate}}
\put(3,54){$(\alpha',\beta')$}
\caption{\label{fig:FigCommDia}%
An (infinitesimal) hypersurface deformation with parameters 
$(\alpha_1,\beta_1)$ that maps $\Sigma\mapsto\Sigma_1$,
followed by one with parameters $(\alpha_2,\beta_2)$ that 
maps $\Sigma_1\mapsto\Sigma_{12}$ differs by one with 
parameters $(\alpha',\beta')$ given by (\ref{eq:X(ab)CommValues})
from that in which the maps with the same parameters are 
composed in the opposite order.}
\end{figure}

Alternatively, decomposing (\ref{eq:XofV}) into normal and tangential 
components with respect to the leaves of the embedding at which the 
tangent-vector field to $\EmbSpace$ is evaluated, yields an 
embedding-dependent parametrisation of $X(V)$ in terms of 
$(\alpha,\beta)$,
\begin{equation}
\label{eq:X-alphabeta}
X(\alpha,\beta)=
\int_\Sigma d^3x
\Bigl(\alpha(x)n^\mu[y](x)+\beta^m(x)\partial_m y^\mu(x)
\Bigr)\,\frac{\delta}{\delta y^\mu(x)}\,,
\end{equation}
where $y$ in square brackets indicates the functional dependence 
of $n$ on the embedding. The functional derivatives of $n$ with 
respect to $y$ can be computed (see the Appendix of 
\cite{Teitelboim:1973}) and the commutator of deformation generators 
then follows to be, 
\begin{equation}
\label{eq:X(ab)Comm}
\bigl[X(\alpha_1,\beta_1)\,,\,X(\alpha_2,\beta_2)\bigr] 
=\,-\,X(\alpha',\beta')\,,
\end{equation}
where 
\begin{subequations}
\label{eq:X(ab)CommValues}
\begin{eqnarray}
\label{eq:X(ab)Comm-b}
&\alpha' &\,=
\beta_1(\alpha_2)-\beta_2(\alpha_1)\,,\\
\label{eq:X(ab)Comm-c}
&\beta' &\,=
[\beta_1,\beta_2]+\sigma\alpha_1\,\mathrm{grad}_h(\alpha_2)
     -\sigma\alpha_2\mathrm{grad}_h\,(\alpha_1)\,.
\end{eqnarray}
\end{subequations}
Here we left open whether spacetime $M$ is Lorentzian ($\sigma=1$) 
or Euclidean ($\sigma=-1$), just in order to keep track how the 
signature of spacetime, $(-\sigma,+,+,+)$, enters. Note that 
the $h$-dependent gradient field for the scalar function $\alpha$ is 
given by $\mathrm{grad}_h(\alpha)=(h^{ab}\partial_b\alpha)\partial_a$. 
The geometric idea behind (\ref{eq:X(ab)CommValues}) is summarised 
in Figure\,\ref{fig:FigCommDia}. 

\subsection{Hamiltonian geometrodynamics}
The idea of Hamiltonian Geometrodynamics is to realise these 
relations in terms of a Hamiltonian system on the phase space of 
physical fields. The most simple case is that where the latter 
merely include the spatial metric $h$ on $\Sigma$, so that the 
phase space is the cotangent bundle $T^*\Riem$ over $\Riem$.
One then seeks a correspondence 
\begin{equation}
\label{eq:PhaseSpaceDist}
(\alpha,\beta)\mapsto\bigl(
H(\alpha,\beta)\,:\,T^*\Riem\rightarrow\reals\bigr)\,,
\end{equation}
where
\begin{equation}
\label{eq:NormalDefCorr}
H(\alpha,\beta)[h,\pi]:=
\int_\Sigma d^3x\bigl(
\alpha(x)\Hcal[h,\pi](x)+h_{ab}(x)\beta^a(x)\Dcal^b[h,\pi](x)\bigr)\,,
\end{equation}
with integrands $\Hcal[h,\pi](x)$ and $\Dcal^b[h,\pi](x)$ yet to 
be determined. $H$ should be regarded as distribution (here the test 
functions are $\alpha$ and $\beta^a$) with values in real-valued 
functions on $T^*\Riem$. Now, the essential requirement is that 
the Poisson brackets between the $H(\alpha,\beta)$ are, up to a 
minus sign,\footnote{Due to the standard convention that the 
Hamiltonian action being defined as a \emph{left} action, whereas 
the Lie bracket on a group is defined by the commutator of 
left-invariant vector fields which generate \emph{right} 
translations.} as in (\ref{eq:X(ab)CommValues}): 
\begin{equation}
\label{eq:H(ab)Comm}
\bigl\{H(\alpha_1,\beta_1)\,,\,H(\alpha_2,\beta_2)\bigr\}
=H(\alpha',\beta')\,.
\end{equation}

Once the distribution $H$ satisfying (\ref{eq:H(ab)Comm}) has been 
found, we can turn around the arguments given above and recover the 
action of the Lie algebra of four-dimensional diffeomorphism on the 
extended phase space including embedding 
variables~\cite{Isham.Kuchar:1985b}. That such an extension is 
indeed necessary has been shown in~\cite{Pons:2003}, where 
obstructions against the implementation of the action of the Lie 
algebra of four-dimensional diffeomorphisms have been identified 
in case the dynamical fields include non-scalar ones.

\subsection{Why constraints}
From this follows a remarkable uniqueness result. Before stating 
it with all its hypotheses, we show why the constraints 
$\Hcal[h,\pi]=0$ and $\Dcal^b[h,\pi]=0$ must be imposed. 

Consider the set of smooth real-valued functions on
phase space, $F:T^*\Riem\rightarrow\reals$. They are 
acted upon by all $H(\alpha,\beta)$ via Poisson bracketing:
$F\mapsto\bigl\{F,H(\alpha,\beta)\bigr\}$. This defines a
map from $(\alpha,\beta)$ into the derivations of phase-space
functions. We require this map to also respect the 
commutation relation (\ref{eq:H(ab)Comm}), that is, we require 
\begin{equation}
\label{eq:Der(ab)Comm-1}
\bigl\{\bigl\{F,H(\alpha_1,\beta_1)\bigr\},H(\alpha_2,\beta_2)\bigr\}-
\bigl\{\bigl\{F,H(\alpha_2,\beta_2)\bigr\},H(\alpha_1,\beta_1)\bigr\}
=\bigl\{F,H\bigr\}(\alpha',\beta')\,.
\end{equation}
The subtle point to be observed here is the following: Up to now  
the parameters $(\alpha_1,\beta_1)$ and $(\alpha_2,\beta_2)$ were 
considered as given functions of $x\in\Sigma$, independent of the 
fields $h(x)$ and $\pi(x)$, i.e. independent of the point of phase 
space. However, from (\ref{eq:X(ab)Comm-c}) we see that $\beta'(x)$ 
does depend on $h(x)$. This dependence may not give rise to extra terms 
$\propto\{F,\alpha'\}$ in the Poisson bracket, for, otherwise, the
extra terms would prevent the map 
$(\alpha,\beta)\mapsto\bigl\{-,H(\alpha,\beta)\bigr\}$
from being a homomorphism from the algebraic structure of 
hypersurface deformations into the derivations of phase-space 
functions. This is necessary in order to interpret 
$\bigl\{-,H(\alpha,\beta)\bigr\}$ as a generator (on phase-space 
functions) of a \emph{spacetime} evolution corresponding to a normal 
lapse $\alpha$ and tangential shift $\beta$. In other words, the 
evolution of observables from an initial hypersurface $\Sigma_i$ 
to a final hypersurface $\Sigma_f$ must be independent of the 
intermediate foliation (`integrability' or `path independence'
~\cite{Teitelboim:1973,Hojman.etal:1973,Hojman.etal:1976}).
Therefore we placed the parameters $(\alpha',\beta')$ outside the 
Poisson bracket on the right-hand side of (\ref{eq:Der(ab)Comm-1}), 
to indicate that no differentiation with respect to $h,\pi$ should 
act on them. 

To see that this requirement implies the constraints, rewrite 
the left-hand side of (\ref{eq:Der(ab)Comm-1}) in the form
\begin{equation}
\label{eq:Der(ab)Comm-2}
\begin{split}
& \bigl\{\bigl\{F,H(\alpha_1,\beta_1)\bigr\},H(\alpha_2,\beta_2)\bigr\}-
\bigl\{\bigl\{F,H(\alpha_2,\beta_2)\bigr\},H(\alpha_1,\beta_1)\bigr\}\\
&\quad=\,\bigl\{F,\bigl\{H(\alpha_1,\beta_1),H(\alpha_2,\beta_2)\bigr\}\bigr\}\\
&\quad=\,\bigl\{F,H(\alpha',\beta')\bigr\}\\
&\quad=\,\bigl\{F,H\bigr\}(\alpha',\beta')
  +H\bigl(\{F,\alpha'\}\,,\,\{F,\beta'\}\bigr)\,,
\end{split}
\end{equation}
where the first equality follows from the Jacobi identity, the second 
from (\ref{eq:H(ab)Comm}), and the third from the Leibniz rule. 
Hence the requirement (\ref{eq:Der(ab)Comm-1}) is equivalent to 
\begin{equation}
\label{eq:Der(ab)Comm-3}
H\bigl(\{F,\alpha'\}\,,\,\{F,\beta'\}\bigr)=0
\end{equation}
for all phase-space functions $F$ to be considered and all 
$\alpha',\beta'$ of the form (\ref{eq:X(ab)CommValues}). 
Since only $\beta'$ depends on phase space, more precisely on $h$, 
this implies the vanishing of the phase-space functions 
$H\bigl(0,\{F,\beta'\}\bigr)$ for all $F$ and all $\beta'$ 
of the form (\ref{eq:X(ab)Comm-c}). This can be shown to 
imply $H(0,\beta)=0$, i.e. $\Dcal[h,\pi]=0$. Now, 
in turn, for this to be preserved under all evolutions we need 
$\bigl\{H(\alpha,\tilde \beta),H(0,\beta)\bigr\}=0$, and hence 
in particular $\bigl\{H(\alpha,0),H(0,\beta)\bigr\}=0$ for all 
$\alpha,\beta$, which implies $H(\alpha,0)=0$, i.e. $\Hcal[h,\pi]=0$. 
So we see that the constraints indeed follow. 

Sometimes the constraints $H(\alpha,\beta)=0$ are split into 
the \emph{Hamiltonian (or scalar) constraints}, $H(\alpha,0)=0$, and 
the \emph{diffeomorphisms (or vector) constraints}, $H(0,\beta)=0$. 
The relations (\ref{eq:H(ab)Comm}) with (\ref{eq:X(ab)CommValues})
then show that the vector constraints form a 
Lie-subalgebra which, because of $\{H(0,\beta),H(\alpha,0)\}
=H\bigl(\beta(\alpha),0\bigr)\ne H(0,\beta')$, is not an 
ideal. This means that the Hamiltonian vector fields for the 
scalar constraints are not tangent to the surface of 
vanishing  vector constraints, except where it intersects the 
surface of vanishing scalar constraints. This implies that the 
scalar constraints do not act on the solution space for the 
vector constraints, so that one simply cannot first reduce the 
vector constraints and then, on the solutions of that, search  
for solutions to the scalar constraints. Also, it is sometimes 
argued that the scalar constraints should \emph{not} be regarded 
as generators of gauge transformations but rather as generators 
of physically meaningful motions whose effect is to change the 
physical state in a fashion that is, in principle, observable. 
See \cite{Kuchar:1993} and also \cite{Barbour.Foster:2008} and 
Sect.\,2.3 of Claus Kiefer's contribution for a recent revival 
of that discussion. However, it seems inconsistent to me to 
simultaneously assume 
 1)~physical states to always satisfy the scalar constraints and
 2)~physical observables to exist which do not Poisson commute
with the scalar constraints: The Hamiltonian vector field 
corresponding to such an `observable' will not be tangent to 
the surface of vanishing scalar constraints and hence will 
transform physical to unphysical states upon being actually 
measured.

\subsection{Uniqueness of Einstein's geometrodynamics}
It is sometimes stated that the relations (\ref{eq:H(ab)Comm})
together with (\ref{eq:X(ab)CommValues}) determine the 
function $H(\alpha,\beta):T^*\Riem\rightarrow\reals$, i.e. 
the integrands $\Hcal[h,\pi]$ and $\Dcal[h,\pi]$, uniquely 
up to two free parameters, which may be identified with the 
gravitational and the cosmological constants. This is a 
mathematical overstatement if read literally, since the 
result can only be shown if certain additional assumptions 
are made concerning the action of $H(\alpha,\beta)$ on the 
basic variables $h$ and $\pi$. 

The first such assumption concerns the intended (`semantic' or
`physical') meaning of $H(0,\beta)$, namely that the action
of $H(0,\beta)\}$ on $h$ or $\pi$ is that of an infinitesimal 
spatial diffeomorphism of $\Sigma$. Hence it should be the 
spatial Lie derivative, $L_\beta$, applied to $h$ or $\pi$. 
It then follows from the general Hamiltonian theory that 
$H(0,\beta)$ is given by the \emph{momentum map} that maps the 
vector field $\beta$ (viewed as element of the Lie algebra 
of the group of spatial diffeomorphisms) into the function 
on phase space given by the contraction of the momentum 
with the $\beta$-induced vector field $h\rightarrow L_\beta h$ 
on $\Riem$:
\begin{equation}
\label{eq:MomentiumMap}
H(0,\beta)=\int_\Sigma d^3x\,
\pi^{ab}(L_\beta h)_{ab}
=-2\int_\Sigma d^3x (\nabla_a\pi^{ab})h_{bc}\beta^c\,.
\end{equation}
Comparison with (\ref{eq:NormalDefCorr}) yields
\begin{equation}
\label{eq:DiffConstrFunct}
\Dcal^b[h,\pi]=-2\nabla_a\pi^{ab}\,.
\end{equation}

The second assumption concerns the intended (`semantic' or
`physical') meaning of $H(\alpha,0)$, namely that 
$\{-,H(\alpha,0)\}$ acting on $h$ or $\pi$ is that of an 
infinitesimal `timelike' diffeomorphism of $M$ normal to 
the leaves $\Emb_t(\Sigma)$. If $M$ were given, it is 
easy to prove that we would have $L_{\alpha n}h=2\alpha\,K$, 
where $n$ is the timelike field of normals to the leaves 
$\Emb_t(\Sigma)$ and $K$ is their extrinsic curvature.
Hence one requires 
\begin{equation}
\label{eq:Anticip}
\{h,H(\alpha,0)\}=2\alpha\,K\,.
\end{equation}
Note that both sides are symmetric covariant tensor fields over
$\Sigma$. The important fact to be observed here is that $\alpha$
appears without differentiation. This means that $H(\alpha,0)$ 
is an ultralocal functional of $\pi$, which is further assumed to be 
a polynomial. (Note that we do not assume any relation between 
$\pi$ and $K$ at this point).

Quite generally, we wish to stress the importance of such 'semantic' 
assumptions concerning the intended meanings of symmetry operations 
when it comes to `derivations' of physical laws from `symmetry principles'.
Such derivations often suffer from the same sort of overstatement 
that tends to give the impression that the mere requirement that 
some group $G$ acts as symmetries alone distinguishes some dynamical 
laws from others. Often, however, additional assumptions are made 
that severely restrict the form in which $G$ is allowed to act. 
For example, in field theory, the requirement of locality often 
enters decisively, like in the statement that Maxwell's vacuum 
equations are Poincar\'e- but not Galilei invariant. In fact, 
without locality the Galilei group, too, is a symmetry group of
vacuum electrodynamics~\cite{Fushchich.Shtelen:1991}. Coming back 
to the case at hand, I do not know of a uniqueness result that 
does not make the assumptions concerning the spacetime interpretation 
of the generators $H(\alpha,\beta)$. Compare also the related 
discussion in~\cite{Samuel:2000b,Samuel:2000a}. 

The uniqueness result for Einstein's equation, which in its 
space-time form is spelled out in Lovelock's theorem 
\cite{Lovelock:1972} already mentioned above, now takes the 
following form in Geometrodynamics~\cite{Kuchar:1973}:
\begin{theorem}
\label{thm:Uniqueness}
In four spacetime dimensions (Lorentzian for $\sigma=1$, 
Euclidean for $\sigma=-1$), the most general functional 
(\ref{eq:NormalDefCorr}) satisfying (\ref{eq:H(ab)Comm})
with (\ref{eq:X(ab)CommValues}), subject to the conditions 
discussed above, is given by (\ref{eq:DiffConstrFunct}) 
and the two-parameter $(\kappa,\Lambda)$ family
\begin{equation}
\label{eq:HamConstrFunct}
\Hcal[h,\pi]=\sigma\,(2\kappa)\, G_{ab\,cd}\pi^{ab}\pi^{cd}
-(2\kappa)^{-1}\sqrt{\det(h)}\bigl(R(h)-\Lambda\big)\,,
\end{equation}
where
\begin{equation}
\label{eq:DeWittMetricMomenta}
 G_{ab\,cd}=\tfrac{1}{2\sqrt{\det(h)}}\bigl(
h_{ac}h_{bd}+h_{ad}h_{bc}-\tfrac{1}{2}h_{ab}h_{cd}\bigr)\,,
\end{equation}
and $R(h)$ is the Ricci scalar of $(h,\Sigma)$. Note that 
(\ref{eq:DeWittMetricMomenta}) is just the ``contravariant version'' 
of the metric (\ref{eq:WDWmetric-c}) for $\lambda=1$, i.e., 
$G_{ab\,nm}G^{nm\,cd}=\frac{1}{2}(\delta^c_a\delta^d_b+\delta^d_a\delta^c_b)$. 
\end{theorem}

The Hamiltonian evolution so obtained is precisely that of 
General Relativity (without matter) with gravitational 
constant $\kappa=8\pi G/c^4$ and cosmological constant 
$\Lambda$. The proof of the theorem is given in \cite{Kuchar:1973},
which improves on earlier versions \cite{Teitelboim:1973,Hojman.etal:1976} 
in that the latter assumes in addition that $\Hcal[h,\pi]$ be 
an even function of $\pi$, corresponding to the requirement 
of time reversibility of the generated evolution.  This was 
overcome in \cite{Kuchar:1973} by the clever move to write 
the condition set by $\{H(\alpha_1,0),H(\alpha_2,0)\}=H(0,\beta')$ 
(the right-hand side being already known) on $H(\alpha,0)$ in 
terms of the corresponding Lagrangian 
functional $L$, which is then immediately seen to turn into a 
condition which is \emph{linear} in $L$, so that terms with 
even powers in velocity decouple form those with odd powers. 
However, a small topological subtlety remains that is neglected 
in all these references and which potentially introduces a little 
more ambiguity that that encoded in the two parameters $\kappa$ and 
$\Lambda$, though its significance is more in the quantum theory. 
To see this recall that we can always perform a canonical 
transformation of the form
\begin{equation}
\label{eq:CanTrans}
\pi\mapsto \pi':=\pi+\Theta
\end{equation}
where $\Theta$ is a closed one-form on $\Riem$. The latter condition 
ensures that all Poisson brackets remain the same if $\pi$ is 
replaced with $\pi'$. Since $\Riem$ is an open positive convex cone 
in a vector space and hence contractible, it is immediate that 
$\Theta=d\theta$ for some function $\theta:\Riem\rightarrow\reals$.
However, $\pi$ and $\pi'$ must satisfy the diffeomorphism constraint, 
which is equivalent to saying that the kernel of $\pi$ (considered 
as one-form on $\Riem$) contains the vertical vector fields, which 
implies that $\Theta$, too, must annihilate all $V_\xi$ so that 
$\theta$ is constant on each connected component of the $\DiffF$ 
orbit in $\Riem$. But unless the $\DiffF$ orbits in $\Riem$ are 
connected, this does not mean that $\theta$ is the pull back of a
function on Superspace, as assumed in \cite{Kuchar:1973}. We can 
only conclude that $\Theta$ is the pull back of a closed but not 
necessarily exact one-form on Superspace. Hence there is an 
analogue of the Bohm-Aharonov-like  ambiguity that one always 
encounters if the configuration space is not simply connected. 
Whether this is the case depends in a determinate fashion on the 
topology of $\Sigma$: One has, due to the contractibility of $\Riem$, 
\begin{equation}
\label{eq:SuperspaceHomotopyGroups}
\pi_n\bigl(\Riem/\DiffF\bigr)\cong\pi_{n-1}\bigl(\DiffF\bigr) \qquad
(n\geq 1)\,.
\end{equation}
For $n=1$ the right hand side is
\begin{equation}
\label{eq:SuperspaceFundGroup}
\pi_{0}\bigl(\DiffF\bigr):=\DiffF/\DiffFConn=:\MCGF
\end{equation}
where $\DiffFConn$ is the identity component of $\DiffF$ and where we 
introduced the name $\MCGF$ (Mapping-Class Group for Frame 
fixing diffeomorphisms) for the quotient group of components.

In view of the uniqueness result above, one might wonder what goes wrong 
when using (the contravariant version of) the metric $G_\lambda$ for 
$\lambda\ne 1$ in (\ref{eq:DeWittMetricMomenta}). The answer is that 
it would spoil (\ref{eq:H(ab)Comm}). More precisely, it would 
contradict $\{H(\alpha_1,0)\,,\,H(\alpha_2,0)\}
=H(0,\alpha_1\nabla\alpha_2-\alpha_2\nabla\alpha_1)$ due to an 
extra term $\propto (h_{ab}\pi^{ab})^2$ in $\Hcal[h,\pi]$, unless 
the additional constraint $h_{ab}\pi^{ab}=0$ were imposed, which 
is equivalent to $\tr_h(K)=0$ and hence to the condition that only 
maximal slices are allowed~\cite{Giulini:1995c}. 
But this is clearly unacceptable (cf. Sect.\,6 
of~\cite{Barbour.etal:2002}).

As a final comment about uniqueness of representations 
of~(\ref{eq:H(ab)Comm}) we mention the apparently larger 
ambiguity---labelled by an additional $\complex$-valued 
parameter, the \emph{Barbero-Immirzi parameter}---that one gets 
if one uses connection variables rather than metric variables 
(cf. \cite{Barbero:1995,Immirzi:1997}, 
Sect.\,4.2.2 of \cite{Thiemann:MCQGR},  and Sect.\,4.3.1 of 
\cite{Kiefer:QuantumGravity}). 
However, in this case one does not represent (\ref{eq:H(ab)Comm})
but a semi-direct product of it with the Lie algebra of 
$SU(2)$ gauge transformations, so that after taking the 
quotient with respect to the latter (which form an ideal) our 
original (\ref{eq:H(ab)Comm}) is represented non locally. 
Also, unless the Barbero-Immirzi parameter takes the very 
special values $\pm i$ (for Lorentzian signature; $\pm 1$
for Euclidean signature) the connection variable does not 
admit an interpretation as a space-time gauge field restricted 
to spacelike hypersurfaces (cf. \cite{Immirzi:1997,Samuel:2000a}). 
For example, the holonomy of a spacelike curve $\gamma$ varies 
with the choice of the spacelike hypersurface containing $\gamma$,
which would be impossible if the spatial connection were the 
restriction of a space-time connection~\cite{Samuel:2000a}. 
Accordingly, the dynamics generated by the constraints does 
then not admit the interpretation of being induced by 
appropriately moving a hypersurface through a spacetime with 
fixed geometric structures on it. Consequently, the argument provided 
here for why one should require (\ref{eq:H(ab)Comm}) in the first 
place does, strictly speaking, not seem to apply in case of 
connection variables. It is therefore presently unclear to me 
on what set of assumptions a uniqueness result could be 
based in this case.

\section{Topology of configuration space}
\label{sec:TopologySup}
Much of the global topology of $\SupF$ is encoded in its homotopy 
groups, which, in turn are given by those of $\DiffF$ according to 
(\ref{eq:SuperspaceHomotopyGroups}). Their structures were 
investigated in \cite{Witt:1986b,Giulini:1995a,Giulini:1997a}.
Early references as to their possible relevance in quantum gravity
are \cite{Friedman.Sorkin:1980,Isham:1982,Sorkin:1986,Sorkin:1989}.

We start by remarking that topological invariants of $\SupF$
are also topological invariants of $\Sigma$, which need not be 
homotopy invariant of $\Sigma$ even if they are homotopy invariants 
of $\SupF$. This is, e.g., the case for the mapping-class group of 
homeomorphisms~\cite{McCarty:1963} and hence (in 3 dimensions) also 
for the mapping-class group $\MCGF$. Remarkably, this means 
that we may distinguish homotopy equivalent but non homeomorphic 
3-manifolds by looking at homotopy invariants of their associated 
Superspaces. Examples for this are given by certain types of lens
spaces. First recall the definition of lens spaces $L(p,q)$ in 
3 dimensions: $L(p,q)=S^3/\!\!\sim$, where $(p,q)$ is a pair of 
positive coprime integers with $p>1$, 
$S^3=\{(z_1,z_2)\in\complex^2\mid\vert z_1\vert^2+\vert z_2\vert^2=1\}$, 
and $(z_1,z_2)\sim (z'_1,z'_2)\Leftrightarrow z'_1=\exp(2\pi i/p)z_1$, 
and $z'_2=\exp(2\pi i\,q/p)z_2$. One way to picture them is to 
take a solid ball in $\reals^3$ and identify each point on the 
upper hemisphere with a points on the lower hemisphere after a rotation by 
$2\pi q/p$ about the vertical symmetry axis. (Usually one depicts the ball
in a way in which it is slightly squashed along the vertical axis so 
that the equator develops a sharp edge and the whole body looks like a 
lens; see e.g. Fig.\,in~\cite{Seifert.Threlfall:Topology}.) In this way 
each set of $p$ equidistant points on the equator is identified to a single 
point. The fundamental group of $L(p,q)$ is $\integers_p$, 
independent of $q$, and the higher homotopy groups are those of its
universal cover, $S^3$. Moreover, for connected closed orientable 
3-manifolds the homology and cohomology groups are also determined 
by the fundamental group in an easy fashion: If $A$ denotes the 
operation of abelianisation of a group, $F$ the operation of taking 
the free part of a finitely generated abelian group, then the first 
four (zeroth to third, the only non-trivial ones) homology and 
cohomology groups are respectively given by 
$H_*=(\integers,A\pi_1,FA\pi_1,\integers)$ and 
$H^*=(\integers,FA\pi_1,A\pi_1,\integers)$ respectively. 
Hence, if taken of $L(p,q)$, all these standard invariants are 
sensitive only to $p$. However, it is known that $L(p,q)$ and 
$L(p,q')$ are 
\begin{enumerate}
\item
homotopy equivalent iff $qq'=\pm n^2\,(\mathrm{mod}\,p)$ for some 
integer $n$,  
\item
homeomorphic iff (all four possibilities) 
$q'=\pm q^{\pm 1}\,(\mathrm{mod}\,p)$, and
\item
orientation-preserving homeomorphic iff 
$q'=q^{\pm 1}\,(\mathrm{mod}\,p)$.
\end{enumerate}  
The first statement is Theorem\,10 in \cite{Whitehead:1941} and the 
second and third statement follow, e.g., from the like combinatorial 
classification of lens spaces \cite{Reidemeister:1935} together with 
the validity of the `Hauptvermutung' (the equivalence of the 
combinatorial and topological classifications) in 
3~dimensions~\cite{Moise:1952}. So, for example, $L(15,1)$ is 
homotopy equivalent but not homeomorphic to $L(15,4)$. On the other 
hand, it is known that 
the mapping-class group $\MCGF$ for $L(p,q)$ is 
$\integers\times\integers$ if $q^2=1\,(\mathrm{mod}\,p)$ with 
$q\not =\pm 1\,(\mathrm{mod}\,p)$, which applies to $p=15$ and 
$q=4$, and that in the remaining cases for $p>2$ it is just 
$\integers$ (see Table\,IV on p.\,591 of~\cite{Witt:1986b}). 
Hence $\MCGF\cong\integers\times\integers$ for $\Sigma=L(15,4)$ 
and $\MCGF\cong\integers$ for $\Sigma=L(15,1)$, 
even though $L(15,1)$ and $L(15,4)$ are homotopy equivalent!

\begin{figure}[ht]
\centering
\includegraphics[width=\linewidth]{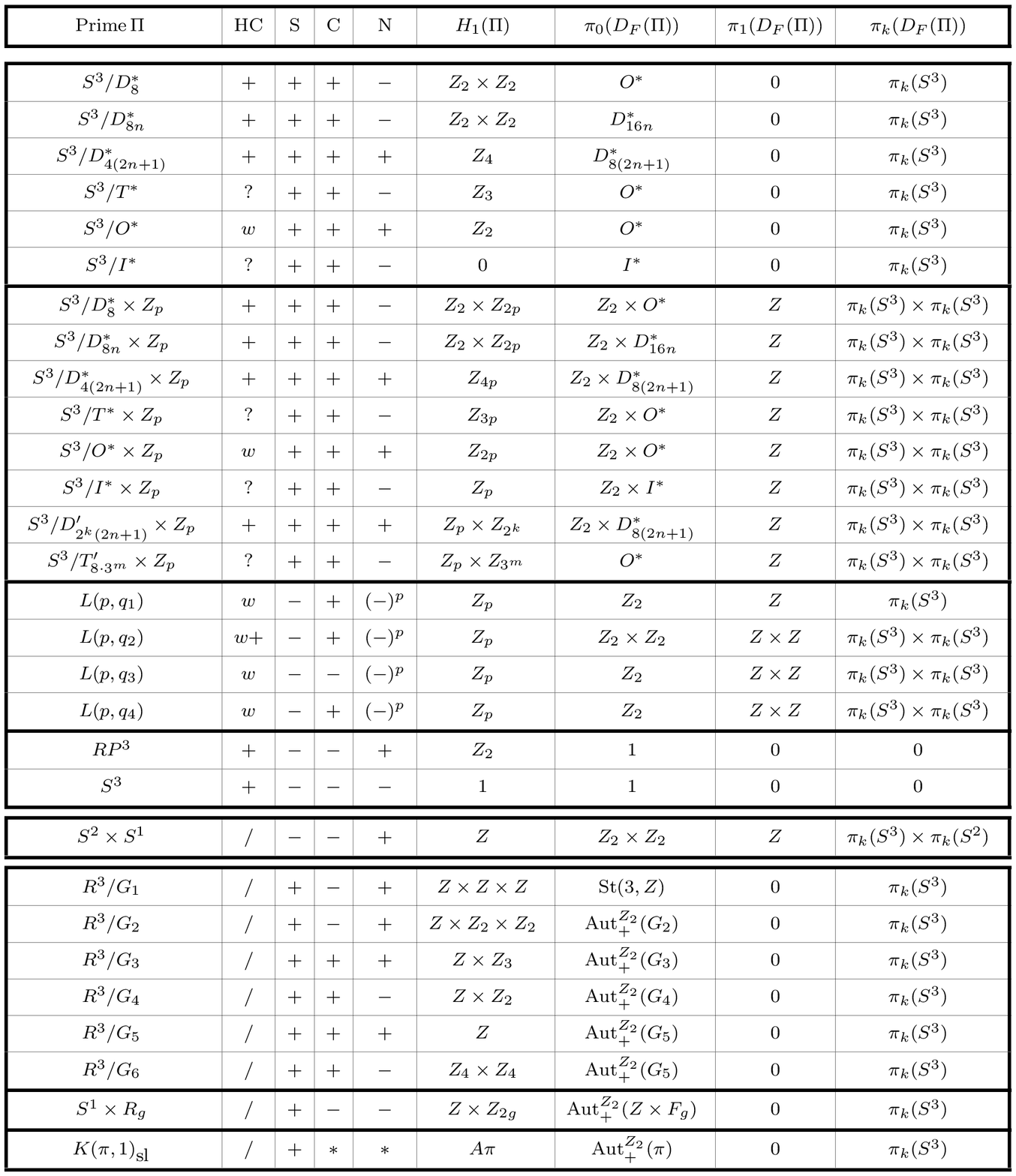}
\caption{\label{Fig:3ManifoldsTable}
This table, taken from \cite{Giulini:1994a}, lists various 
properties of certain prime 3-manifolds. The manifolds are 
grouped into those of finite fundamental group, which are 
of the form $S^3/G$, the exceptional one, $S^1\times S^2$, which 
is prime but not irreducible ($\pi_2(S^1\times S^2)=\mathbb{Z}$),
those six which can carry a flat metric and which are of the 
form $\mathbb{R}^3/G$, and so-called Haken manifolds 
(sufficiently large $K(\pi,1)$ primes). For the lens spaces 
$q_1$ stands for $q=\pm 1$, $q_2$ for $q\ne\pm 1$ and $q^2=1$,
$q_3$ for $q^2=-1$, and $q_4$ for the remaining cases, where 
all equalities are taken $\mathrm{mod}\,p$. The 3rd and 4th column 
list spinoriality and chirality, the last three columns the 
homotopy groups of their corresponding superspace. We refer to 
 \cite{Giulini:1994a} for the meanings of the other columns.} 
\end{figure}

Quite generally it turns out that Superspace stores much 
information about the topology of the underlying 3-manifold 
$\Sigma$. This can be seen from the table in 
Figure\,\ref{Fig:3ManifoldsTable}, which we reproduced form 
\cite{Giulini:1994a}, and where properties of certain 
prime manifolds (see below for an explanation of `prime') are 
listed. There is one interesting observation from that list which 
we shall mention right away: From gauge theories it is known that 
there is a relation between topological invariants of the classical 
configuration space and certain features of the corresponding 
quantum-field theory~\cite{Jackiw:LesHouches1983}, in particular 
the emergence of certain anomalies which represent non-trivial 
topological invariants~\cite{Alvarez-GaumeGinsparg:1985}. 
By analogy one could conjecture similar relations to hold quantum 
gravity. An interesting question is then whether there are 
preferred manifolds $\Sigma$ for which all these invariants are trivial. 
From those represented on the table there is indeed a unique pair 
of manifolds for which this is the case, namely the 3-sphere and 
the 3-dimensional real projective space. 
To understand more of the information collected in the table we 
have to say more about general 3-manifolds.  

Of particular interest is the fundamental group of Superspace.  
Experience with ordinary quantum mechanics (cf.~\cite{Giulini:1995b}
and references therein) already suggests that its classes of 
inequivalent irreducible unitary representations correspond to a 
superselection structure which here might serve as fingerprint 
of the topology of $\Sigma$ in the quantum theory. The sectors 
might, e.g., correspond to various statistics (in the presence of 
diffeomorphic primes) that preserve or violate a 
naively expected spin-statistics correlation %
\cite{Aneziris.etal:1989a,Aneziris.etal:1989b,Dowker.Sorkin:1998,Dowker.Sorkin:2000}
(see also below).

\subsection{General three-manifolds and specific examples}
\label{sec:GeneralThreeManifolds}

\begin{figure}[ht]
\centering
\includegraphics[width=0.8\linewidth]{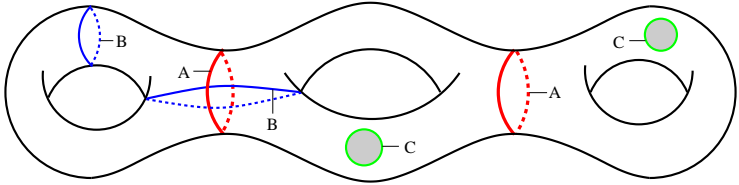}
\caption{\label{fig:FigRiemannSurface}
A Riemann surface of genus\,3 with three pairs of embedded
1-spheres (circles) of type A, B, and C. Type A is essential
and splitting, type B is essential but not splitting, and type 
C is splitting but not essential. Any third essential and 
splitting 1-sphere can be continuously deformed via embeddings
into one of the two drawn here.}
\end{figure}

The way to understand general 3-manifolds is by cutting them 
along certain embedded two manifolds so that the remaining pieces 
are simpler in an appropriate sense. Here we shall only consider 
those simplifications that are achieved by cutting along embedded 
2-spheres. (Further decompositions by cutting along 2-tori 
provide further simplifications, but these are not directly relevant 
here.) The 2-spheres should be `essential' and `splitting'.
An essential 2-sphere is one which does not bound a 3-ball and a 
splitting 2-sphere is one whose complement has two (rather than 
just one) connected components. Figure\,\ref{fig:FigRiemannSurface} 
is intended to visualise the analogues of these notions in 
two dimensions.
Given a closed 3-manifold $\Sigma$, consider the following 
process: Cut it along an essential splitting 2-sphere and 
cap off the 2-sphere boundary of each remaining component 
by a 3-disk. Now repeat the process for each of the remaining 
closed 3-manifolds. This process stops after a finite number 
of steps~\cite{Kneser:1929} where the resulting components 
are uniquely determined up to diffeomorphisms (orientation 
preserving if oriented manifolds are considered) and 
permutation~\cite{Milnor:1962}; see \cite{Hatcher:3-manifolds} 
for a lucid discussion. The process stops at that stage at 
which none of the remaining components, $\Pi_1,\cdots,\Pi_n$, 
allows for essential splitting 2-spheres, i.e. at which 
each $\Pi_i$ is a prime manifold. A 3-manifold is called 
\emph{prime} if each embedded 2-sphere either bounds a 3-disc 
or does not split; it is called \emph{irreducible} if each 
embedded 2-sphere bounds a 3-disc. In the latter case the 
second homotopy group, $\pi_2$, must be trivial, since, if 
it were not, the so-called sphere theorem 
(see, e.g., \cite{Hatcher:3-manifolds}) ensured the 
existence of a non-trivial element of $\pi_2$ which could 
be represented by an \emph{embedded} 2-sphere. Conversely, it 
follows from the validity of the Poincar\'e conjecture that 
a trivial $\pi_2$ implies irreducibility. Hence irreducibility 
is equivalent to a trivial $\pi_2$. There is precisely one 
non-irreducible prime 3-manifold, and that is the handle 
$S^1\times S^2$. Hence a 3-manifold is prime iff it is 
either a handle or if its $\pi_2$ is trivial. 

\begin{figure}[ht]
\begin{minipage}[c]{0.35\linewidth}
\centering\includegraphics[width=0.85\linewidth]{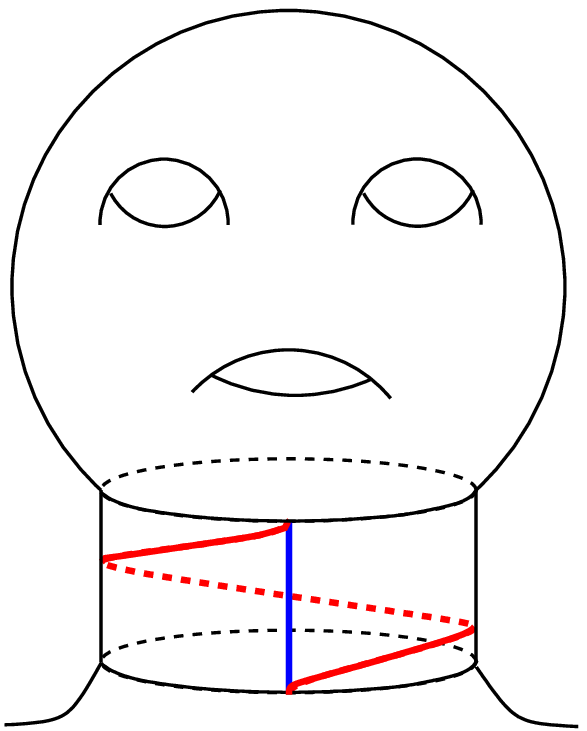}
\end{minipage}
\hfill
\begin{minipage}[c]{0.50\linewidth}
\centering\includegraphics[width=0.85\linewidth]{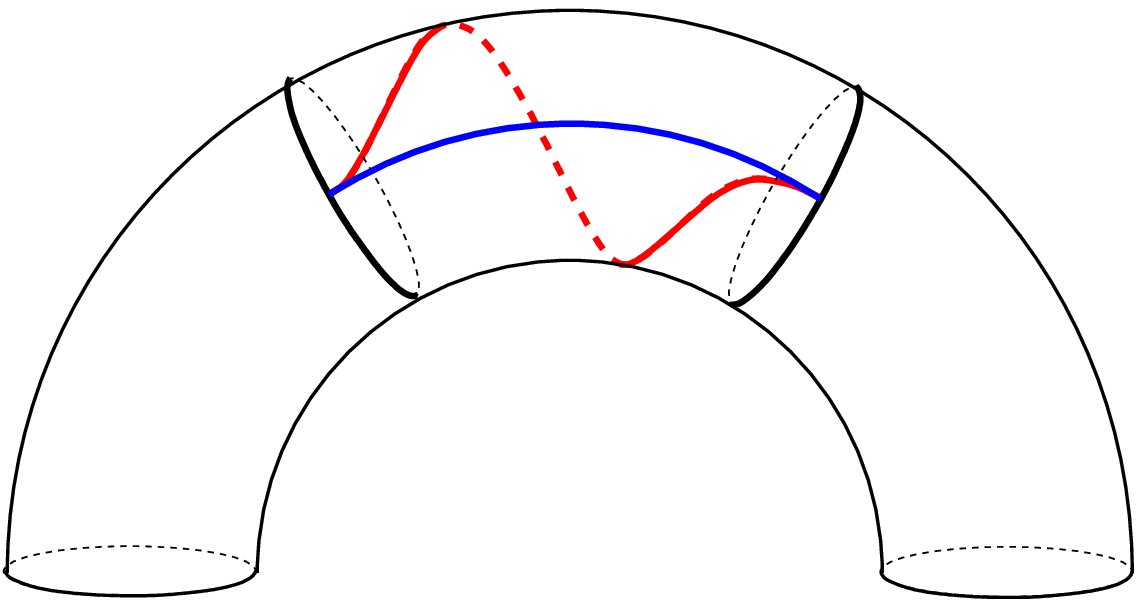}
\end{minipage}
\put(-50,12){\small $S_1$}
\put(-130,12){\small $S_2$}
\put(-290,50){\large $\Pi$}
\put(-290,-21){$S_1$}
\put(-290,-64){$S_2$}
\caption{\label{fig:FigDehnTwists}%
Both pictures show rotations parallel to spheres $S_1$ and $S_2$:
On the left, a rotation of a prime manifold in a connected sum 
parallel to the connecting sphere, on the right a rotation 
parallel to two meridian spheres in a `handle' $S^1\times S^2$.
The support of the diffeomorphism is on the cylinder bound by 
$S_1$ and $S_2$. In either case its effect is depicted by the 
two curves connecting the two spheres. The two-dimensional
representation given here is deceptive insofar, as in two 
dimensions the original and the mapped curves are not homotopic
(keeping their endpoints fixed), due to the one-sphere not being 
simply connected, whereas they are in three (and higher) 
higher dimensions they are due to the higher-dimensional spheres 
being simply connected.  
}
\end{figure}

Given a general 3-manifold $\Sigma$ as connected sum of primes
$\Pi_1,\cdots,\Pi_n$, there is a general method to establish 
$\MCGF$ in terms of the individual mapping-class groups of the 
primes. The strategy is to look at the effect of elements in 
$\MCGF$ on the fundamental group of $\Sigma$. As $\Sigma$ is 
the connected sum of primes, and as connected sums in $d$
dimensions are taken along $d-1$ spheres which are 
simply-connected for $d\geq 3$, the fundamental group of a 
connected sum is the free product of the fundamental groups 
of the primes for $d\geq 3$. The group $\MCGF$ now naturally 
acts as automorphisms of $\pi(\Sigma)$ by simply taking the 
image of a based loop that represents an element in 
$\pi(\Sigma)$ by a based (same basepoint) diffeomorphism that 
represents the class in $\MCGF$. Hence there is a natural map
\begin{equation}
\label{eq:MCGmapstoAut}
d_F:\MCGF\rightarrow\mathrm{Aut}\bigl(\pi_1(\Sigma)\bigr)\,.
\end{equation}
The known presentations\footnote{%
A (finite) presentation of a group is its characterisation in 
terms of (finitely many) generators and (finitely many) relations.}
of automorphism groups of free products in terms of presentations 
of the automorphisms of the individual factors and additional 
generators (basically exchanging isomorphic factors and conjugating 
whole factors by individual elements of others) can now be used 
to establish (finite) presentations of $\MCGF$, provided 
(finite) presentations for all prime factors are 
known.\footnote{This presentation of the automorphism group of
free products is originally due to Fouxe-Rabinovitch 
\cite{Fouxe-Rabinovitch:1940,Fouxe-Rabinovitch:1941}. Modern forms 
with corrections are given in \cite{McCullough.Miller:1986} and 
\cite{Gilbert:1987}} Here I wish to stress that this situation would 
be more complicated if $\Diff$ rather than $\DiffF$ (or at least
the diffeomorphisms fixing a preferred point) had been considered; 
that is, had we not made the transition from (\ref{eq:DefSuperspace})
to (\ref{eq:DefSupF}). Only for $\DiffF$ (or the slightly larger 
group of diffeomorphisms fixing the point) is it generally true 
that the mapping-class group of a prime factor injects into the 
mapping-class group of the connected sum in which it appears. 
For more on this, compare the discussion on p.\,182-3 in~\cite{Giulini:2007a}.
Clearly, one also needs to know which elements are in the kernel 
of the map (\ref{eq:MCGmapstoAut}). This will be commented on below
in connection with Fig~\ref{fig:FigDehnTwists}.

\begin{figure}[ht]
\centering\includegraphics[width=0.64\linewidth]{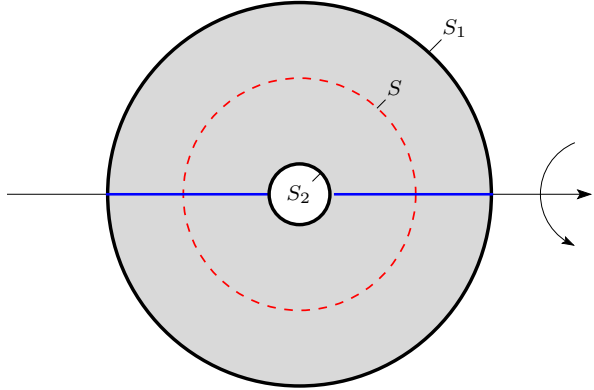}
\put(-57,134){$S_1$}
\put(-116,71){$S_2$}
\put(-78,111){$S$}
\caption{\label{fig:TwoRP3}%
The connected sum of two real projective spaces may be 
visualised by the shaded region obtained from the 
spherical shell that is obtained by rotating the shaded 
annulus about the vertical symmetry axis as indicated.
Antipodal points on the outer 2-sphere boundary $S_1$,
as well as on the inner 2-sphere boundary $S_2$, are 
pairwise identified. This results in the connected sum 
of two $\reals P^3$ along the connecting sphere $S$. 
Due to the antipodal identifications, the two thick 
horizontal segments in the shaded region become a single
loop, showing that the entire space is fibred by circles 
over $\reals P^2$. Remarkably, the handle $S^1\times S^2$, 
which is prime, is a double cover of this reducible manifold.}  
\end{figure}

\begin{figure}[ht]
\begin{minipage}[c]{0.45\linewidth}
\centering\includegraphics[width=0.70\linewidth]{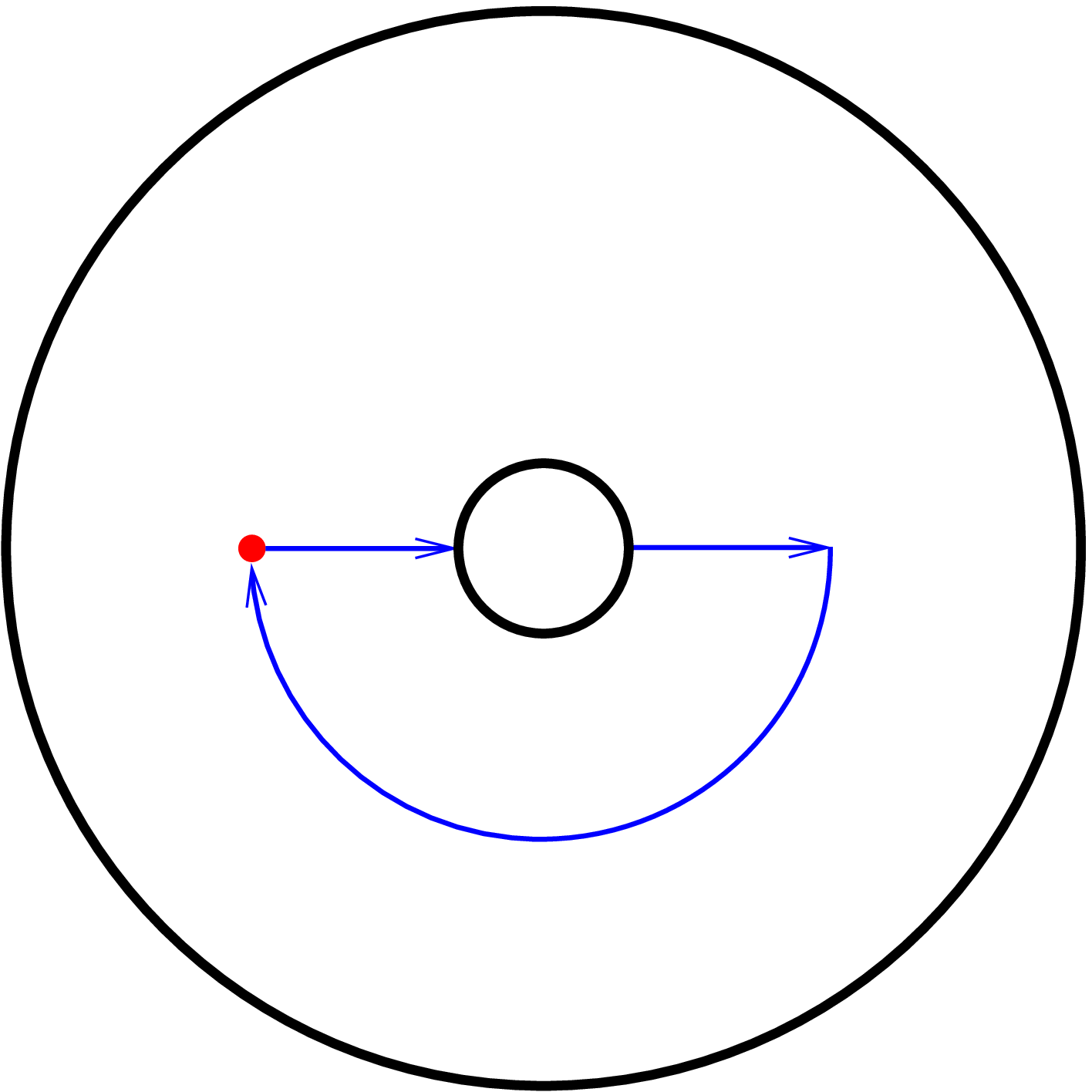}
\put(-58,75){\Large $a$}
\end{minipage}
\hfill
\begin{minipage}[c]{0.45\linewidth}
\centering\includegraphics[width=0.70\linewidth]{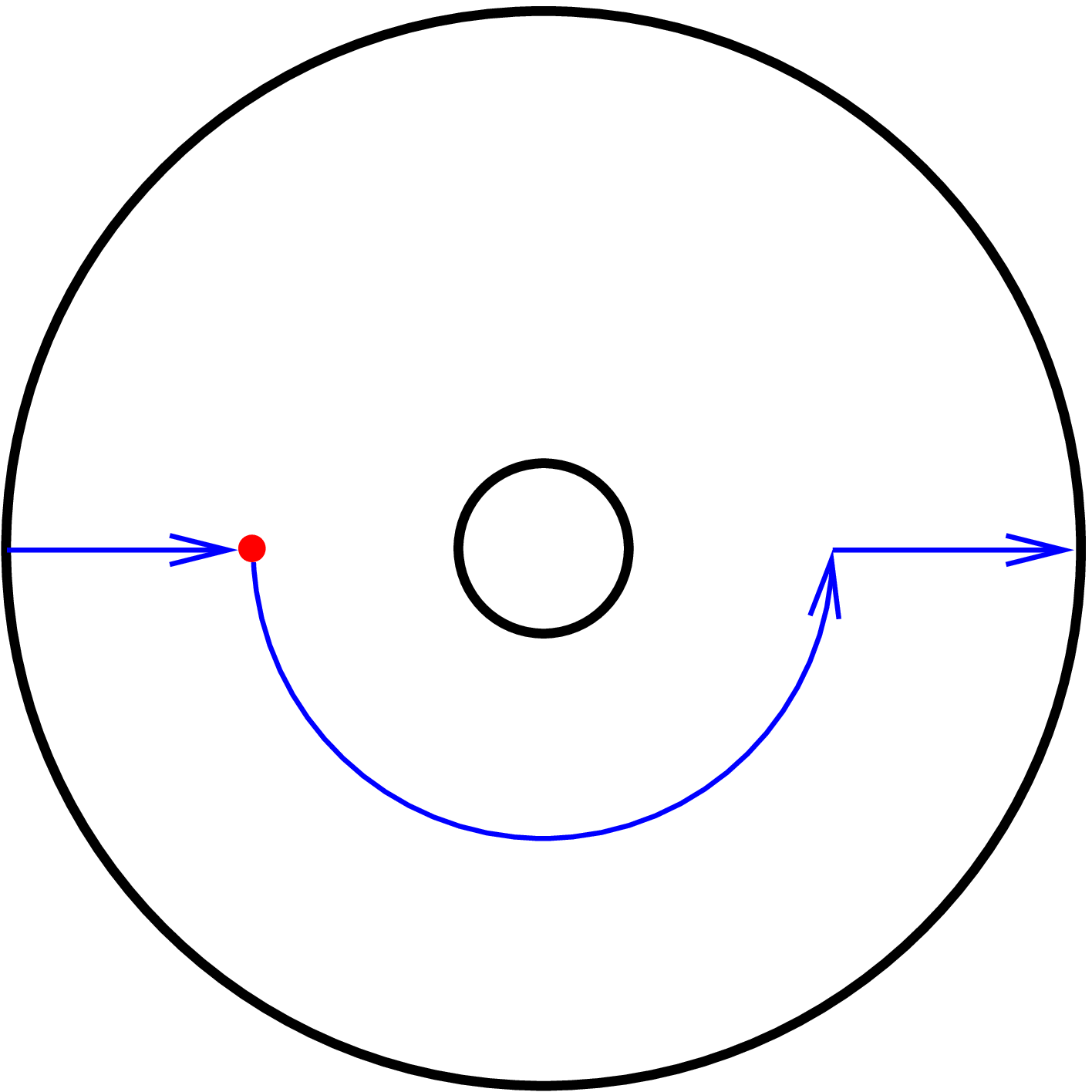}
\put(-58,75){\Large $b$}
\end{minipage}
\caption{\label{fig:GenFundGrTwoRP3}%
The shown loops represent the generators $a$ and $b$ of the fundamental 
group $\integers_2*\integers_2=\langle a,b\mid a^2=b^2=1\rangle$.
The product loop $c=ab$ is then seen to be one of the fibres mentioned  
in the caption of the previous Figure. In terms of $a$ and $c$ we 
have the presentation $\langle a,c\mid a^2=1,\, aca^{-1}=c\rangle$, 
showing the isomorphicity 
$\integers_2*\integers_2\cong\integers_2\ltimes\integers$.}
\end{figure}
\subsection{The connected sum of two real-projective spaces}
In some (in fact many) cases the map $d_F$ is an isomorphism. For example, this is
the case if $\Sigma$ is the connected sum of two $\reals P^3$, so that 
$\pi_1(\Sigma)$ is the free product $\integers_2*\integers_2$, a 
presentation of which is $\langle a,b\mid a^2=b^2=1\rangle$. For the 
automorphisms we have $\mathrm{Aut}(\integers_2*\integers_2)\cong
\integers_2*\integers_2=\langle E,S\mid E^2=S^2=1\rangle$, where 
$E:(a,b)\rightarrow (b,a)$ and $S:(a,b)\rightarrow (a,aba^{-1})$. 
In this sense the infinite discrete group $\integers_2*\integers_2$
is a quotient of the automorphism group of Superspace $\SupF$
for $\Sigma$ being the connected sum of two real projective spaces.   
It is therefore of interest to study its unitary irreducible 
representations. This can be done directly in a rather elementary 
fashion, or more systematically by a simple application of the 
method of induced representations (Mackey theory) using the 
isomorphicity $\integers_2*\integers_2\cong\integers_2\ltimes\integers$
(cf.~caption to Fig.\,\ref{fig:GenFundGrTwoRP3}).
The result is that, apart form the obvious four one-dimensional ones,
given by $(E,S)\mapsto (\pm\id,\pm\id)$, there is a continuous set
of mutually inequivalent two-dimensional ones, given by 
\begin{equation}
\label{eq:MCGRepTwoRP3s}
E\mapsto
\left(
\begin{array}{lr}
1&0\\ 0&-1
\end{array}\right)\,,\qquad
S\mapsto
\left(
\begin{array}{lr}
\cos\tau & \sin\tau\\ 
\sin\tau & -\cos\tau\\
\end{array}\right)\,,\quad
\tau\in(0,\pi)\,.
\end{equation}
Already in this most simple example of a non-trivial connected sum 
we have an interesting structure in which the two `statistics 
sectors' corresponding to the irreducible representations of the 
permutation subgroup (here just given by the $\integers_2$ subgroup 
generated by $E$) get mixed by $S$, where the `mixing angle' $\tau$
uniquely characterises the representation. This behaviour can also 
be studied in more complicated 
examples~\cite{Giulini:1994a}\cite{SorkinSurya:1998}. 
For a more geometric understanding of the maps representing $E$ 
and $S$, see~\cite{Giulini:2007a}.

\subsection{Spinoriality}
I also wish to mention one very surprising observation
that was made by Rafael Sorkin and John Friedman in 
1980~\cite{Friedman.Sorkin:1980} and which has to do with 
the physical interpretation of the elements in the kernel of the 
map (\ref{eq:MCGmapstoAut}), leading to the conclusion that 
pure (i.e. without matter) quantum gravity should already 
contain states with half-integer angular momenta. The reason 
being a purely topological one, depending entirely on the 
topology of $\Sigma$. In fact, given the right topology of 
$\Sigma$, its one-point decompactification used in the context 
of asymptotically flat initial data  will describe an 
isolated system whose asymptotic (at spacelike infinity) 
symmetry group is not the ordinary Poincar\'e 
group~\cite{Beig.Murchadha:1987} but rather its double 
(= universal) cover. This gives an intriguing answer to 
Wheeler's quest to find a natural place for spin\,1/2 in 
Einstein's standard geometrodynamics 
(cf. \cite{Misner.Thorne.Wheeler:Gravitation} Box\,44.3).
 
I briefly recall that after introducing the concept of a `Geon' 
(`gravitational-electromagnetic entity') in 1955~\cite{Wheeler:1955}, 
and inspired by the observation that electric charge (in the 
sense of non-vanishing flux integrals of $\star F$ over 
closed 2-dimensional surfaces) could be realised in Einstein-Maxwell 
theory without sources (`charge without charge'), Wheeler and 
collaborators turned to the Einstein-Weyl 
theory~\cite{Brill.Wheeler:1957} and tried to find a `neutrino 
analog of electric charge'~\cite{Klauder.Wheeler:1957}. Though 
this last attempt failed, the programme of `matter as geometry' 
in the context of geometrodynamics, as outlined in the 
contributions to the anthology~\cite{Wheeler:Geometrodynamics},
survived in Wheeler's thinking well into the 1980s~\cite{Wheeler:1982}.  

Back to the `spin without spin' topologies, the elements of the 
kernel of (\ref{eq:MCGmapstoAut}) can be pictured as rotation parallel 
to certain spheres, as depicted in Figure\,\ref{fig:FigDehnTwists}. 
(In many---and possibly all---cases the group generated by 
such maps actually exhaust the kernel; compare Theorem.\,1.5 in 
\cite{McCullough:1990} and footnote\,21 in \cite{Giulini:2007a}). 
The point we wish to focus on here is that for \emph{some} prime 
manifolds the diffeomorphism depicted on the left in 
Figure\,\ref{fig:FigDehnTwists} is indeed not in the identity 
component of all diffeomorphisms that fix a frame exterior to 
the outer ($S_2$) 2-sphere. Such manifolds are called 
\emph{spinorial}. For each prime it is known whether it is 
spinorial or not, and the easy-to-state but hard-to-prove
result is, that the only non-spinorial manifolds\footnote{
We remind the reader that `manifold' here stands for 
`3-dimensional closed orientable manifold'.} are the lens 
spaces $L(p,q)$, the handle $S^1\times S^2$, and connected 
sums amongst them. That these manifolds are not spinorial 
is, in fact, very easy to visualise. Hence, given the proof 
of the `only' part and of the fact that a connected sum is 
spinorial iff it contains at least one spinorial prime, one 
may summarise the situation by saying that that the only 
non-spinorial manifolds are the `obvious' ones. 

Even though being a generic property in the sense just stated, 
spinoriality is generally hard to prove in dimensions three or 
greater. This is in marked contrast to two dimensions, where the 
corresponding transformation shown in the left picture of 
Figure\,\ref{fig:FigDehnTwists} acts non trivially on the 
fundamental group. Indeed, consider a base point outside 
(below) $S_2$ in the left picture in Figure\,\ref{fig:FigDehnTwists}, 
then the rotation acts by conjugating each of the $2g$ generators 
$(a_1\cdots a_g,b_1\cdots b_2)$ that $\Pi$ adds to $\pi_1(\Sigma)$ by 
the element $\prod_{i=1}^g a_ib_ia^{-1}_ib^{-1}_i$, which is 
non-trivial in $\pi_1(\Sigma)$ if other primes exist or otherwise 
if the point with the fixed frame is removed (as one may do, due 
to the restriction to diffeomorphisms fixing that point).  

\begin{figure}[ht]
\centering\includegraphics[width=0.4\linewidth]{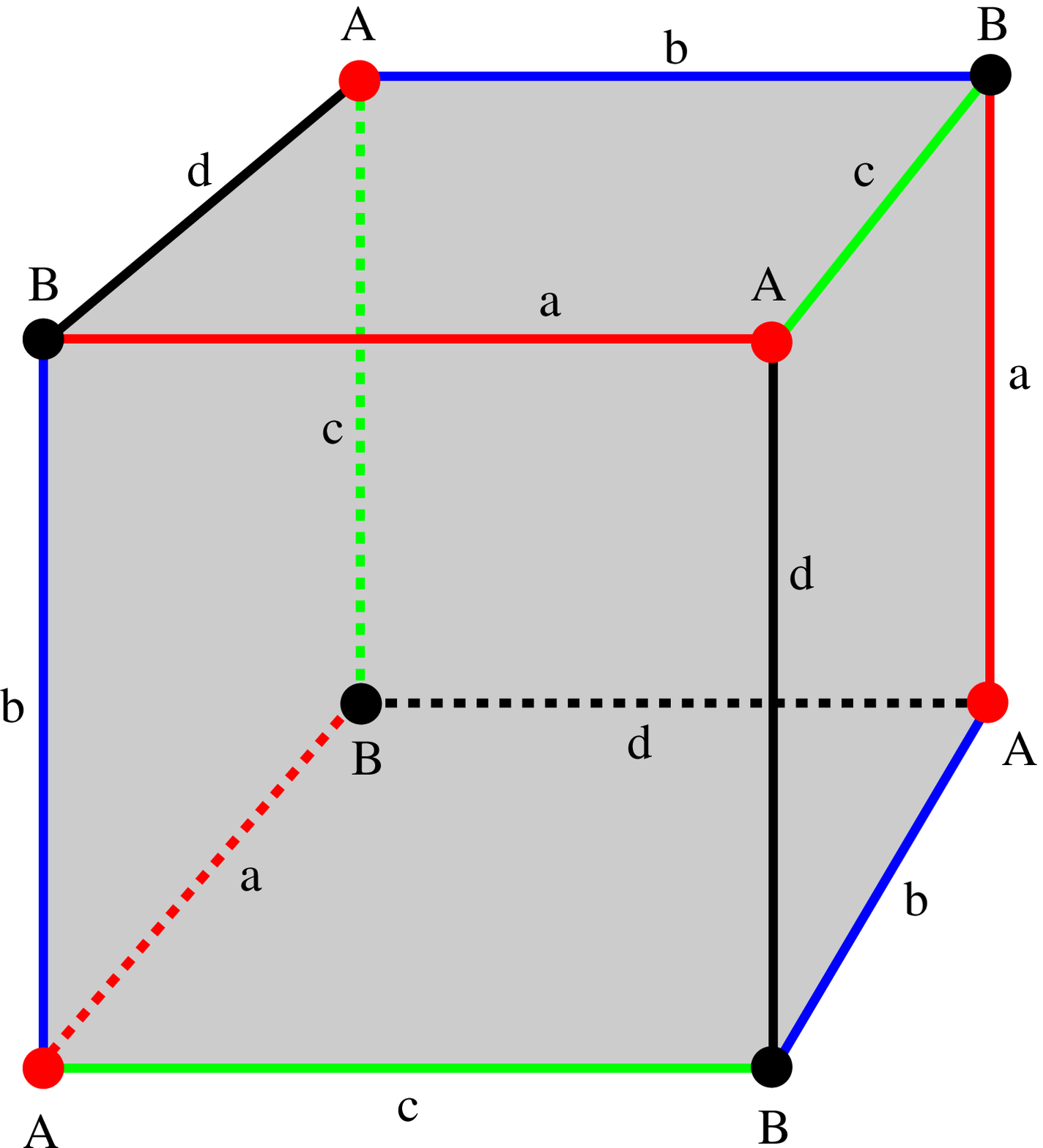}
\caption{\label{fig:Q-Space}%
Fundamental domain for the space $S^3/D^*_8$. Opposite faces are 
identified after a right-handed screw motion with $90$-degree 
rotation, as indicated by the coinciding labels for the edges
and vertices. The corresponding space with a left-handed 
identification is not orientation-preserving diffeomorphic to 
this one.}
\end{figure}

An example of a spinorial manifold is the spherical space form 
$S^3/D^*_8$, where $S^3$ is thought of as the sphere of unit 
quaternions and $D_8^*$ is the subgroup in the group of unit 
quaternions given by the eight elements $\{\pm 1,\pm i, \pm j, \pm k\}$. 
The coset space $S^3/D^*_8$ may be visualised as solid cube 
whose opposite faces are identified after a $90$-degree 
rotation by either a right- or a left-handed screw motion;
see Figure\,\ref{fig:Q-Space}. Drawings of fundamental domains in 
form of (partially truncated) solid polyhedra with suitable boundary
identifications for spaces $S^3/G$ are given in~\cite{DuVal:Homographies}. 
Let us take the basepoint $\infty\in S^3/D^*_8$ to be the centre 
of the cube in Figure\,\ref{fig:Q-Space}. The two generators, $a$ 
and $b$, of the fundamental group
\begin{equation}
\label{eq:FundGrPresQSpace}
\pi_1(S^3/D^*_8)\cong D^*_8=\langle a,b\mid a^2=b^2=(ab)^2\rangle
\end{equation}
are then represented by two of the three oriented straight 
segments connecting the midpoints of opposite faces. The third 
corresponds to the product $ab$. A rotation of the cube about 
its centre by any element of its crystallographic symmetry group
defines a diffeomorphism of $S^3/D^*_8$ fixing $\infty$ 
since it is compatible with the boundary identification. 
It is not in the identity component of $\infty$-fixing 
diffeomorphisms since it obviously acts non-trivially on 
the generators of the fundamental group. Clearly, each such 
rigid rotation may be modified in an arbitrarily small 
neighbourhood of $\infty$ so as to also fix the tangent 
space at this point. That $S^3/D^*_8$ is spinorial means 
that in going from the point-fixing to the frame-fixing 
diffeomorphisms one acquires more diffeomorphisms not 
connected to the identity. More precisely, the mapping-class
group of frame-fixing diffeomorphisms is a $\integers_2$-extension 
of the mapping-class group of merely point-fixing 
diffeomorphisms. The generator of this extending $\integers_2$ 
is a full $360$-degree rotation parallel to two small
concentric spheres centred at $\infty$. In this way, the 
spinoriality of $S^3/D^*_8$ extends the crystallographic 
symmetry group $O\subset SO(3)$ of the cube to its double 
cover $O^*\subset SU(2)$ and one finally gets
\begin{equation}
\label{eq:MCGofQSpace}
\mathrm{MCG}_{\mathrm{F}}(\Pi)\cong O^*\quad\mathrm{for}\quad \Pi=S^3/D^*_8\,.
\end{equation}
This is precisely what one finds at the intersection of the 
2nd row and 7th column of the table in 
Figure\,\ref{Fig:3ManifoldsTable}, with corresponding results for 
the other spherical space forms. 

Coming back to the previous example of the connected sum of two 
(or more \cite{Giulini:1994a}) real projective spaces, we can make 
the following observation: First of all, real projective 3-space is   
a non-spinorial prime. This is obvious once one visualises it as
a solid 3-ball whose 2-sphere boundary points are pairwise 
identified in an antipodal fashion, since this identification is 
compatible with a rigid rotation. A full rotation about the, say,
centre-point of the ball may therefore be continuously undone by 
a rigid rotation outside a small ball about the centre, suitably 
`bumped off' towards the centre. Second, as we have seen above, 
the irreducible representations of the mapping-class group of 
the connected sum contains both statistics sectors independently 
for one-dimensional representations and in a mixed form for the 
continuum of two-dimensional irreducible representations. This 
already shows~\cite{Aneziris.etal:1989b} that there is no 
general kinematical spin-statistics relation as in other 
non-linear theories~\cite{Finkelstein.Rubinstein:1968,Sorkin:1988}.   
Such a relation may at best be re-introduced for some manifolds 
by restricting the way in which states are constructed, e.g., 
via the sum-over-histories approach~\cite{Dowker.Sorkin:1998}.

\subsection{Chirality}
There is one last aspect about diffeomorphisms that can be explained
in terms of Figure\,\ref{fig:Q-Space}. As stated in the caption of 
this figure, there are two versions of this space: one where the 
identification of opposite faces is done via a $90$-degree 
right-handed screw motion and one where one uses a left-handed screw 
motion. These spaces are not related by an orientation preserving 
diffeomorphism. This is equivalent to saying that, say, the first 
of these spaces has no orientation-reversing self-diffeomorphism. 
Manifolds for which this is the case are called \emph{chiral}. 
There are no examples in two dimensions. To see this, just consider 
the usual picture of a Riemannian genus $g$ surface embedded into 
$\reals^3$ and map it onto itself by a reflection at any of its 
planes of symmetry. So chiral manifolds start to exist in 3 
dimensions and continue to do so in all higher dimensions, as 
was just recently shown~\cite{Muellner:2008}.  If one tries to 
reflect the cube in Figure\,\ref{fig:Q-Space} at one of its 
symmetry planes one finds that this is incompatible with the 
boundary identifications, that is, pairs of identified points are 
not mapped to pairs of identified points. Hence this reflection 
simply does not define a map of the quotient space. This clearly does 
not prove the nonexistence of orientation reversing maps, since 
there could be others than these obvious candidates. 

In fact, following an idea of proof in \cite{Witt:1986b}, the 
chirality of $S^G/D^*_8$ (and others of the form $S^3/G$) can be 
reduced to that of $L(4,1)$. That the latter is chiral follows 
from the following argument: Above we have already stated that 
$L(p,q)$ and $L(p,q')$ are orientation preserving diffeomorphic iff 
$q'=q^{\pm 1}\,(\mathrm{mod}\,p)$. Since taking the mirror 
image in $\reals^3$ of the lens representing $L(p,q)$ gives 
the lens representing $L(p,-q)$, $L(p,q)$ admits an orientation 
reversing diffeomorphism iff $L(p,q)$ and $L(p,-q)$ are orientation 
preserving diffeomorphic. But, as just stated, this is the case iff 
$-q=q^{\pm 1}\,(\mathrm{mod}\,p)$, i.e. if either $p=2,\,q=1$ 
(recall that $p$ and $q$ must be coprime) or 
$q^2=-1\,(\mathrm{mod}\,p)$.\footnote{Remarkably, this result 
was already stated in footnote\,1 of p.\,256 of \cite{Kneser:1929}.
An early published proof is that in Sect.\,77 of 
\cite{Seifert.Threlfall:Topology}}  
Hence, in particular, all $L(p,1)$ are chiral. Now, $G=D_8^*$ has 
three subgroups isomorphic to 
$\integers_4$, the fundamental group of $L(4,1)$, namely the ones 
generated by $i$, $j$, and $k$. They are normal so that we have a 
regular covering $L(4,1)\displaystyle{\mathop{\rightarrow}^p} S^3/G$. 
Now suppose $f: S^3/G\rightarrow S^3/G$ were orientation 
reversing, i.e. a diffeomorphism with $\deg(f)=-1$. Consider the diagram 
\begin{equation}
\label{eq:DiagChirality}
\bfig
\qtriangle|amr|/.>`>`>/<970,600>[L(4{,}1)`L(4{,}1)`S^3/G;\tilde f`f\circ p`p]
\btriangle|lmb|/>`>`>/<970,600>[L(4{,}1)`S^3/G`S^3/G;p`f\circ p`f]
\efig
\end{equation}
If the lift $\tilde f$ existed we would immediately get a 
contradiction since from commutativity of (\ref{eq:DiagChirality})
we would get $\deg(\tilde f)\cdot\deg(p)=\deg(p)\cdot\deg(f)$ 
and hence $\deg(\tilde f)=-1$, which contradicts chirality 
of $L(4,1)$. Now, according to the theory of covering spaces 
the lift $\tilde f$ of $f\circ p$ exists iff the image of 
$\pi_1\bigl(L(4,1)\bigr)$ under $(f\circ p)_*$, which is a 
subgroup $\integers_4\subset D_8^*$, is conjugate to the 
image of $\pi_1\bigl(L(4,1)\bigr)$ under $p_*$. This need not 
be the case, however, as different subgroups $\integers_4$ in 
$D_8^*$ are normal and hence never conjugate (here we 
deviate from the argument in \cite{Witt:1986b} which seems 
incorrect). However, by composing a given orientation 
reversing $f$ with an orientation preserving diffeomorphism 
that undoes the $\integers_4$ subgroup permutation introduced by 
$f$, we can always create a new orientation reversing diffeomorphism 
that does not permute the $\integers_4$ subgroups. That new 
orientation reversing diffeomorphism---call it again $f$---now 
indeed has a lift $\tilde f$, so that finally we arrive at the 
contradiction envisaged above. 

As prime manifolds, the two versions of a chiral prime corresponding to 
the two different orientations count as different. This means the following:
Two connected sums which differ only insofar as a particular chiral 
prime enters with different orientations are not orientation-preserving 
diffeomorphic; they are not diffeomorphic at all if the complement 
of the selected chiral prime also chiral (i.e. iff another chiral 
prime exists). For example, the connected sum of two oriented 
$S^3/D_8^*$ is not diffeomorphic to the connected sum of $S^3/D_8^*$
with $\overline{S^3/D_8^*}$, where the overbar indicates the opposite
orientation. Note that the latter case also leads to two 
non-homeomorphic 3-manifolds whose classic invariants (homotopy, 
homology, cohomology) coincide. This provides an example of a 
topological feature of $\Sigma$ that is not encoded into the structure 
of $\SupF$.

\section{Summary and outlook}
\label{sec:SummaryOutlook}
Superspace is for geometrodynamics what gauge-orbit space is 
for non-abelian gauge theories, though Superspace has generally 
a much richer topological and metric structure. Its topological 
structure encodes much of the topology of the underlying 
3-manifold and one may conjecture that some of its topological 
invariants bear the same relation to anomalies and sectorial structure 
as in the case of non-abelian gauge theories. Recent progress in 
3-manifold theory now allows to make more complete statements, in particular 
concerning the fundamental groups of Superspaces associated to 
more complicated 3-manifolds. Its metric structure is piecewise 
nice but also suffers from singularities, corresponding to 
signature changes, whose physical significance is unclear. 
Even for simple 3-manifolds, like the 3-sphere, there are 
regions in superspace where the metric is strictly Lorentzian
(just one negative signature and infinitely many pluses), like 
at the round 3-sphere used in the FLRW cosmological models, so 
that the Wheeler-DeWitt equation becomes strictly hyperbolic,  
but there are also regions with infinitely many negative signs
in the signature. 

Note that the cotangent bundle over Superspace is not the fully 
reduced phase space for matter-free General Relativity. It only 
takes account of the vector constraints and leaves the scalar
constraint unreduced. However, under certain conditions, the 
scalar constraints can be solved by the `conformal method'
which leaves only the conformal equivalence class of 
3-dimensional geometries as physical configurations. In those 
cases the fully reduced phase space is the cotangent bundle over 
conformal superspace, whose analog to (\ref{eq:DefSuperspace})
is given by replacing $\Diff$ by the semi-direct product 
$\Conf\rtimes\Diff$, where $\Conf$ is the abelian group of 
conformal rescalings that acts on $\Riem$ via $(f,h)\mapsto f\,h$
(pointwise multiplication), where $f:\Sigma\rightarrow\reals_+$. 
The right action of $(f,\phi)\in\Conf\rtimes\Diff$ on 
$h\in\Riem$ is then given by by $R_{(f,\phi)}(h)=f\phi^*h$, so that, 
using $R_{(f_2,\phi_2)}R_{(f_1,\phi_1)}=R_{(f_1,\phi_1)(f_2,\phi_2)}$, 
the semi-direct product structure is seen to be 
$(f_1,\phi_1)(f_2,\phi_2)=\bigl(f_2(f_1\circ\phi_2),\phi_1\circ\phi_2\bigr)$.
Note that because of $(f_1f_2)\circ\phi=(f_1\circ\phi)(f_2\circ\phi)$
$\Diff$ indeed acts as automorphisms of $\Conf$. Conformal superspace 
and extended conformal superspace would then, in analogy to 
(\ref{eq:DefSuperspace}) and (\ref{eq:DefSupF}), be defined as
$\ConfSup:=\Riem/\Conf\rtimes\Diff$ and $\ConfSupF:=\Riem/\Conf\rtimes\DiffF$
respectively. The first definition was used in \cite{Fischer.Moncrief:1996} 
as applied to manifolds with zero degree of symmetry 
(cf. footnote\,\ref{foot:DegSym}). In any case, since $\Conf$ is 
contractible, the topologies of $\Conf\rtimes\Diff$ and 
$\Conf\rtimes\DiffF$ are those of $\Diff$ and $\DiffF$ which also 
transcend to the quotient spaces analogously to 
(\ref{eq:SuperspaceHomotopyGroups}) whenever the groups act freely. 
In the first case this is essentially achieved by restricting to 
manifolds of vanishing degree of symmetry, whereas in the second 
case this follows almost as before, with the sole exception being 
$(S^3,h)$ with $h$ conformal to the round metric.%
\footnote{\label{foot:ConfIsom} Let $\mathcal{CI}(\Sigma,h):=%
\{\phi\in\Diff\mid\phi^*h=fh,\,f:\Sigma\rightarrow\reals_+\}$ 
be the group of conformal isometries. For compact $\Sigma$ it is 
known to be compact except iff $\Sigma=S^3$ and $h$ conformal 
to the round metric~\cite{Lelong-Ferrand:1971}. Hence, for 
$\Sigma\ne S^3$, we can average $h$ over the compact group 
$\mathcal{CI}(\Sigma,h)$ and obtain a new Riemannian metric 
$h'$ in the conformal equivalence class of $h$ for which 
$\mathcal{CI}(\Sigma,h)$ acts as proper isometries. Therefore, 
by the argument presented in Sect.\ref{sec:Superspace}, it 
cannot contain non-trivial elements fixing a frame.} 
Hence the topological results obtained before also apply 
to this case. In contrast, the geometry for conformal superspace 
differs insofar from that discussed above as the conformal modes that 
formed the negative directions of the Wheeler-DeWitt metric (cf. 
(\ref{eq:WDWmetric-a}) are now absent. The horizontal subspaces 
(orthogonal to the orbits of $\Conf\rtimes\DiffF$) are now given 
by the transverse and traceless  (rather than just obeying 
(\ref{eq:HorizontalityCond})) symmetric two-tensors. In that sense 
the geometry of conformal superspace, if defined as before by some 
ultralocal bilinear form on $\Riem$, is manifestly 
positive (due to the absence of trace terms) and hence less 
pathological than the superspace metric discussed above. 
It might seem that its physical significance is less clear, 
as there is now no constraint left that may be said to induce this 
particular geometry; see however~\cite{Barbour.Murchadha:1999}. 

Whether it is a realistic hope to understand superspace and conformal 
superspace (its cotangent bundle being the space of solutions to 
Einstein's equations) well enough to actually gain a sufficiently complete 
understanding of its automorphism group is hard to say. An interesting 
strategy lies in the attempt to understand the solution space directly 
in a group- (or Lie algebra-) theoretic fashion in terms of a quotient 
$G_\infty/H_\infty$, where $G_\infty$ is an infinite dimensional 
group (Lie algebra) that (locally) acts transitively on the space of 
solutions and $H_\infty$ is a suitable subgroup (algebra), usually 
the fixed-point set of an involutive automorphism of $G$. The basis 
for the hope that this might work in general is the fact that it 
works for the subset of stationary and axially symmetric solutions, 
where $G^\infty$ is the Geroch Group; cf.~\cite{Breitenlohner.Maison:1987}.
The idea for generalisation, even to $d=11$ supergravity, is expressed in 
\cite{Nicolai:1999} and further developed in~\cite{Damour.etal:2007}.   

\bigskip
\noindent
\textbf{Acknowledgements} I thank Hermann Nicolai and Stefan 
Theisen for inviting me to this stimulating 405th WE-Heraeus-Seminar 
on Quantum Gravity and for giving me the opportunity to contribute 
this paper. I also thank Hermann Nicolai for pointing out 
\cite{Nicolai:1999} and Ulrich Pinkall for pointing out 
\cite{Lelong-Ferrand:1971}. 

\bibliographystyle{plain}
\bibliography{RELATIVITY,HIST-PHIL-SCI,MATH,QM} 
\end{document}